\documentclass[prd,onecolumn,groupedaddress, nofootinbib, 11pt]{revtex4}  
\usepackage{graphicx}
\usepackage{epstopdf}
\usepackage{amsmath}
\usepackage{amsfonts}
\usepackage{amssymb}
\usepackage{appendix}
\usepackage{mathtools}
\usepackage{comment}
\usepackage{bbold}
\usepackage{color}
\usepackage{slashed}
\usepackage{subfigure}
\usepackage{setspace}
\usepackage{footnote}
\usepackage[activate={true,nocompatibility},final,tracking=true,kerning=true,spacing=true,factor=1100,stretch=10,shrink=10]{microtype}
\usepackage{multirow}
\usepackage[letterpaper, margin=.8in, top=0.85in, bottom=0.85in]{geometry}

\begin{document}

\singlespacing

\hfill NUHEP-TH/15-03

\title{Global Constraints on a Heavy Neutrino} 

\author{Andr\'{e} de Gouv\^{e}a} 
\affiliation{Northwestern University, Department of Physics \& Astronomy, 2145 Sheridan Road, Evanston, IL 60208, USA}

\author{Andrew Kobach} 
\affiliation{Department of Physics, University of California, San Diego, La Jolla, CA 92093, USA}
\affiliation{Northwestern University, Department of Physics \& Astronomy, 2145 Sheridan Road, Evanston, IL 60208, USA}

\date{\today}

\begin{abstract} 
We estimate constraints on the existence of a heavy, mostly sterile neutrino with mass between 10 eV and 1 TeV.  
We improve upon previous analyses by performing a global combination and expanding the experimental inputs to simultaneously include tests for lepton universality, lepton-flavor-violating processes, electroweak precision data, dipole moments, and neutrinoless double beta decay.  Assuming the heavy neutrino and its decay products are invisible to detection, we further include, in a self-consistent manner, constraints from direct kinematic searches, the kinematics of muon decay, cosmology, and neutrino oscillations, in order to estimate constraints on the values of $|U_{e4}|^2$, $|U_{\mu4}|^2$, and $|U_{\tau4}|^2$. 
\end{abstract}

\maketitle

\tableofcontents

\section{Introduction}

Neutrinos are the least understood particles in the standard model (SM).  
While it is known that there are three neutrino flavor eigenstates that participate in the weak interactions~\cite{ALEPH:2005ab}, referred to as active neutrinos,
it is a generic possibility that there are also electroweak singlet states, called sterile neutrinos, and that the three active flavor eigenstates are linear superpositions of more than just three mass eigenstates, 
\begin{equation}
\label{nulin}
\nu_\alpha = \displaystyle \sum_{i=1}^{3+k} U_{\alpha i} \nu_i, \hspace{0.25in} (\alpha = e, \mu, \tau),
\end{equation}
where $\nu_i$ are neutrino mass eigenstates with mass $m_i$, $U_{\alpha i}$ are elements of a unitary matrix, and $k$ is the number of additional neutrinos beyond those present in the SM.
This manuscript focuses on the hypothesis that the SM is augmented by one new neutrino, i.e., $k=1$, and how experimental results can illuminate this possibility. Throughout, we will assume that there is no relation between the different mixing-matrix elements $U_{\alpha i}$, or any relation between $U_{\alpha i}$ and the different neutrino masses $m_i$.

To date, much attention has been paid to a single Majorana sterile neutrino augmenting the SM,  with mass $10\text{ eV} \lesssim m_4 \lesssim 1 \text{ TeV}$, where the model responsible for its decay is identical to the model that dictates its production, i.e., neutrino production and decay are governed uniquely by the weak interactions.  As discussed, for example, in Refs.~\cite{Atre:2009rg, Drewes:2015iva, Vincent:2014rja, Deppisch:2015qwa}, this particular scenario can be constrained, sometimes severely, by direct searches for the decay products of the heavy neutrino.  However, it is not necessary that neutrinos are Majorana or that the decay of the heavy neutrino is mediated only by weak interactions.  Thus, constraints obtained using a particular model for the heavy neutrino decay must be differentiated from constraints on the {\it existence} of a heavy neutrino.   

Here, we make the phenomenological decision that experiments cannot measure the decay of the heavy neutrino.  Put precisely,  $\nu_4$ decays to other particles that are effectively invisible to direct detection, e.g.,  light neutrinos, dark matter, or other unknown light states.  This assumption provides, in some sense, conservative estimates for upper bounds on the matrix elements $|U_{\alpha4}|^2$, since, in principle, stronger constraints on the matrix elements can be achieved if the heavy neutrino decays into visible particles.  Without observables associated with the heavy neutrino decay, constraints on its existence can change dramatically. Furthermore, neutrino-decay assumptions modify qualitatively how bounds from different types of observables are to be combined. For example, we note that some analyses, e.g., Refs.~\cite{Atre:2009rg,Drewes:2015iva, Deppisch:2015qwa},  include both constraints from experimental searches for specific decay products of the heavy neutrino, and constraints that assume that the heavy neutrino does not decay visibly, such as searches for the kinematic signatures of a heavy neutrino in meson decay.  Here, we pay close attention to the assumptions regarding the decay of the heavy neutrino in order to present self-consistent results.  

Our analysis goes beyond just rearranging previously derived constraints on the values of $|U_{e4}|^2$, $|U_{\mu4}|^2$, and $|U_{\tau4}|^2$.  
We include limits from dedicated kinematic searches, but also include our own estimates for limits using tests of lepton universality (charged-lepton decays, pseudoscalar meson decays, $W$-boson decays, etc), lepton-flavor-violating processes ($\mu-e$ conversion, radiative charged-lepton decays, three-body tau decays, etc.), neutrinoless double beta decay, the spectrum of Michel electrons from muon decay, the invisible decay width of the $Z$ boson, and neutrino oscillations. 
We combine all the relevant experimental results via a global $\chi^2$ function in order to estimate simultaneous upper limits on $|U_{e4}|^2$, $|U_{\mu4}|^2$, and $|U_{\tau}|^2$, when 10 eV$\lesssim m_4\lesssim$ 1 TeV.   

This analysis is outlined as follows.  In Section~\ref{modelindepconstraints}, we discuss constraints on a heavy neutrino that are decay independent, e.g., tests of lepton universality, invisible decays of the $Z$ boson, lepton-flavor-violating processes, and neutrinoless double beta decay.  In Section~\ref{invconst}, we interpret other experimental results --  $\beta$ decay, pseudoscalar meson decay, neutrino oscillations, etc. -- as constraints on a heavy neutrino that decays invisibly.  In Section~\ref{combo}, we describe the details of how we combine the experimental constraints, present the resultant simultaneous limits on $|U_{e4}|^2$, $|U_{\mu4}|^2$, and $|U_{\tau4}|^2$,  discuss the results, highlight important differences with other analyses found in the literature, and offer some concluding remarks.

\section{Decay-Independent Constraints on a Heavy Neutrino}
\label{modelindepconstraints}

This section summarizes constraints on the existence of a single heavy neutrino by focusing on observables with no heavy neutrino in the final state. These observables are independent of whatever physics controls the heavy neutrino decay, and the associated constraints are useful not only for our present analysis but also apply whenever a heavy neutrino is produced through only the weak interactions.  These observables include tests of lepton universality, the width of invisible decay of the $Z$ boson, lepton-flavor-violating decays of $\mu$ and $\tau$ leptons, neutrinoless double beta decay, and magnetic and electric dipole moments of charged leptons.

\subsection{Tests of Lepton Universality}
\label{lepunitests}

In the SM, charged-current interactions couple to the three lepton families, $e$, $\mu$, $\tau$, with a universal constant: $g_e=g_\mu=g_\tau$. Such universality can be studied at the percent and subpercent level by measuring the ratios of decay rates of charged leptons, pseudoscalar mesons, and the $W$ boson.  If a heavy neutrino exists, then the measured values of $|g_\mu/g_e|$, $|g_\tau/g_\mu|$, and $|g_\tau/g_e|$ can deviate from unity.  More concretely, if the heavy neutrino is too heavy to be produced in a given decay, one can relate the comparisons between experiment and predictions  in order to estimate limits on the existence of a heavy neutrino, independent of any assumptions regarding how $\nu_4$ decays.

Lepton universality tests have been used to estimate limits on the existence of a heavy neutrino in a number of analyses (see, for example, Refs.~\cite{Abada:2013aba, Antusch:2014woa, Atre:2009rg, Basso:2013jka, Bertoni:2014mva, Endo:2014hza, Smirnov:2006bu, Nardi:1994iv, delAguila:2008pw,Antusch:2015mia,Fernandez-Martinez:2015hxa}).  We revisit these constraints in hopes to offer self-consistent and precise results by performing a global combination of all relevant {\it ratios} of decay rates, using model-independent methods and the most up-to-date experimental measurements and theoretical predictions.\footnote{We choose to utilize only ratios of decay rates so that several experimental and theoretical uncertainties will, at least partially, cancel. To zeroth order, the bounds extracted are independent from changes to the definitions of fundamental parameters in the presence of the heavy neutrino (e.g., $G_F$ and $\sin^2\theta_W$).}

In Table~\ref{leptuniv}, we compile a list of observables that are sensitive to lepton nonuniversality, comparing experimental results and the SM predictions for ratios of decay rates.  Some details regarding the values in Table~\ref{leptuniv} are itemized here:  
\begin{itemize}
\item {\it Flavor-conserving charged-lepton decays.}  The SM expectations for the ratios of $\Gamma(\tau^-\rightarrow\mu^- \overline{\nu}_\mu \nu_\tau)$, $\Gamma(\tau^-\rightarrow e^- \overline{\nu}_e \nu_\tau)$, and $\Gamma(\mu^-\rightarrow e^- \overline{\nu}_e \nu_\mu)$ at tree level are
\begin{eqnarray}
\frac{\Gamma(\tau^-\rightarrow\mu^- \overline{\nu}_\mu \nu_\tau)}{\Gamma(\tau^-\rightarrow e^- \overline{\nu}_e \nu_\tau)} &=& \frac{f\left(m_\mu^2/m_\tau^2\right)}{f\left(m_e^2/m_\tau^2\right)}, \\
 \frac{\Gamma(\tau^-\rightarrow e^- \overline{\nu}_\mu \nu_\tau)}{\Gamma(\mu^-\rightarrow e^- \overline{\nu}_e \nu_\mu)} &=&  \frac{m_\tau^5}{m_\mu^5}  \frac{f\left(m_e^2/m_\tau^2\right)}{f\left(m_e^2/m_\mu^2\right)}.
\end{eqnarray} 
where $f(x) \equiv 1-8x+8x^3 -x^4-12x^2 \log x$.  The values of the SM expectations for these ratios can be found in Table~\ref{leptuniv}.  These predictions are sufficiently precise for our purposes, since the uncertainties associated with the charged-lepton masses and radiative corrections are an order of magnitude or 2 smaller than the experimental precision~\cite{Pich:2013lsa}.  To compare these SM predictions to experimental values, we use the measured value of $\Gamma(\tau\rightarrow\mu\overline{\nu}\nu )/\Gamma(\tau\rightarrow e\overline{\nu}\nu )$ quoted in Ref.~\cite{Agashe:2014kda}.  Because the experimental measurement of $\Gamma(\tau\rightarrow e\overline{\nu}\nu )/\Gamma(\mu\rightarrow e\overline{\nu}\nu )$ is not performed directly, we estimate its measured value as $\Gamma(\tau\rightarrow e\overline{\nu}\nu )/\Gamma(\mu\rightarrow e\overline{\nu}\nu ) \simeq  \tau_\tau^{-1} \tau_\mu Br(\tau\rightarrow e\overline{\nu}\nu )$, where $\tau_\tau$ and $\tau_\mu$ are the measured lifetimes of the tau and the muon, respectively, and $Br(\tau\rightarrow e\overline{\nu}\nu )$ is the measured branching ratio of $\tau\rightarrow e\overline{\nu}\nu$.  The measured values of $\tau_\tau$, $\tau_\mu$, and $Br(\tau\rightarrow e\overline{\nu}\nu )$ are taken from Ref.~\cite{Agashe:2014kda}.

\item {\it $\pi$, $K$, $K_L$, and $D_s$ decays.}  The  state-of-the-art SM predictions for the ratios $\Gamma(\pi\rightarrow e\nu)/\Gamma(\pi\rightarrow \mu \nu)$, $\Gamma(K\rightarrow e\nu)/\Gamma(K\rightarrow \mu \nu)$, $\Gamma(K\rightarrow \pi \mu\nu)/\Gamma(K\rightarrow \pi e \nu)$, $\Gamma(K_L\rightarrow \pi \mu\nu)/\Gamma(K_L\rightarrow \pi e \nu)$, and $\Gamma(D_s\rightarrow \tau \nu)/\Gamma(D_s\rightarrow \mu\nu)$ are taken from Refs.~\cite{Agashe:2014kda,Cirigliano:2007xi,Wanke:2007km}.  
Table~\ref{leptuniv} includes the experimental measurements of these ratios from Ref.~\cite{Agashe:2014kda}.

\item {\it$\bar{B}^0$ decays.} 
As shown in Table~\ref{leptuniv}, we separate measurements of the branching ratios of $\bar{B}^0$ decays into two categories:  the LHCb experiment measurement of the ratio $\Gamma(\bar{B}^0\rightarrow D^{*+}\tau^- \overline{\nu}_\tau)/\Gamma(\bar{B}^0\rightarrow D^{*+}\mu^- \overline{\nu}_\mu)$~\cite{Aaij:2015yra}, and the Belle and BaBar measurements of $\Gamma(\bar{B}^0\rightarrow D^{(*)+}\tau^- \overline{\nu}_\tau)/\Gamma(\bar{B}^0\rightarrow D^{(*)+}\ell^- \overline{\nu}_\mu)$, where $\Gamma(\bar{B}^0\rightarrow D^{(*)+}\ell^- \overline{\nu}_\mu)$ signifies the average of $\ell=e$ and $\mu$~\cite{Huschle:2015rga, Lees:2012xj} (and we combine the results from Belle and BaBar).  This distinction must be made in order to account for how a heavy neutrino would affect the two measurements differently.   These measurements are compared to the SM expectations from Refs.~\cite{Lattice:2015rga, Fajfer:2012vx}. 

\item {\it$W$ Boson Decays.}  The leptonic decays of the $W$ boson can directly test lepton universality at the weak scale.  The ratios $\Gamma(W\rightarrow \mu \nu)/\Gamma(W\rightarrow  e \nu )$ and $\Gamma(W\rightarrow \tau \nu)/\Gamma(W\rightarrow e \nu )$ are predicted in the SM to be approximately unity, up to radiative corrections and corrections due to the mass of the final-state leptons~\cite{Kniehl:2000rb}.  We quote in Table~\ref{leptuniv} the experimental values for $\Gamma(W\rightarrow \mu \nu)/\Gamma(W\rightarrow  e \nu )$ and $\Gamma(W\rightarrow \tau \nu)/\Gamma(W\rightarrow e \nu )$ from Ref.~\cite{Agashe:2014kda}.

\item Ratios such as $\Gamma(\tau \rightarrow \pi  \nu)/\Gamma(\pi \rightarrow \ell \nu)$ and $\Gamma(\tau \rightarrow K  \nu)/\Gamma(K \rightarrow \ell \nu)$ can be mostly predicted by theory, because the dependence on the decay constants $f_\pi$ and $f_K$ cancels in the ratio.  The SM prediction for these ratios is
\begin{equation}
\frac{\Gamma(\tau^- \rightarrow M^- \nu_\tau )}{\Gamma(M^-\rightarrow \ell^- \overline{\nu}_{\ell})} =  \frac{m_\tau^3}{2m_M m_\ell^2} \left( \frac{1-m_M^2/m_\tau^2}{1-m_\ell^2/m_M^2} \right)^2 \left( 1+ \delta^M_\ell\right),
\end{equation}
where $M=\pi$ or $K$, and $\ell = e$ or $\mu$. The radiative corrections $\delta^\pi_\mu$ and $\delta^K_\mu$ have been estimated in Ref.~\cite{Decker:1994ea}. We estimate the prediction for $\Gamma(\tau \rightarrow M  \nu)/\Gamma(M \rightarrow e \nu)$ by multiplying the predicted value of $\Gamma(\tau \rightarrow M  \nu)/\Gamma(M \rightarrow \mu \nu)$ by the value of $\Gamma(M\rightarrow \mu \nu)/\Gamma(M\rightarrow e \nu )$ calculated in Ref.~\cite{Cirigliano:2007xi}.  The experimental values for these ratios quoted in Table~\ref{leptuniv} are taken from Ref.~\cite{Agashe:2014kda}.

\end{itemize} 

\begin{center}
\begin{table}[h]
\footnotesize
\begin{tabular}{| l | c | c | c | } 
\hline
Observable & SM & Observed & $|g_\ell/g_{\ell'}|^2$  \\ \hline \hline
$\Gamma(\tau\rightarrow \mu \nu \overline{\nu})/\Gamma(\tau\rightarrow e \nu \overline{\nu})$ & 0.9726 & $0.9764\pm0.0030$ & $|g_\mu/g_e|^2= 1.0040 \pm 0.0031$  \\ 
$\Gamma(\pi\rightarrow e \nu)/\Gamma(\pi\rightarrow \mu \nu )$ & $1.235 \times 10^{-4}$~\cite{Cirigliano:2007xi} & $(1.230 \pm 0.004)\times10^{-4}$ & $|g_e/g_\mu|^2= 0.9958 \pm 0.0032 $  \\ 
$\Gamma(K\rightarrow e \nu)/\Gamma(K\rightarrow \mu \nu )$ & $2.477\times10^{-5}$~\cite{Cirigliano:2007xi} & $(2.488 \pm 0.010)\times10^{-5}$ & $|g_e/g_\mu|^2= 1.0044 \pm 0.0040 $  \\ 
$\Gamma(K\rightarrow \pi \mu \nu)/\Gamma(K\rightarrow \pi e \nu )$ & $ 0.6591\pm0.0031$~\cite{Wanke:2007km} & $ 0.6608\pm0.0030$ & $|g_\mu/g_e|^2= 1.0026 \pm 0.0065 $ \\
$\Gamma(K_L\rightarrow \pi \mu \nu)/\Gamma(K_L\rightarrow \pi e \nu )$ & $0.6657\pm0.0031$~\cite{Wanke:2007km} & $ 0.6669\pm0.0027$ & $|g_\mu/g_e|^2= 1.0018 \pm 0.0062 $ \\
$\Gamma(W\rightarrow \mu \nu)/\Gamma(W\rightarrow  e \nu )$ & $ 1.000 $~\cite{Kniehl:2000rb} & $0.993\pm0.019$ & $|g_\mu/g_e|^2= 0.993 \pm 0.020$ \\
\hline \hline
$\Gamma(\tau \rightarrow e \nu \overline{\nu})/\Gamma(\mu \rightarrow e \nu \overline{\nu})$ & $1.345\times10^6$ & $(1.349\pm 0.004)\times10^6$ & $|g_\tau/g_\mu|^2=1.003\pm0.003$ \\ 
$\Gamma(\tau \rightarrow \pi  \nu)/\Gamma(\pi \rightarrow \mu \nu)$ & $9771\pm 14$~\cite{Decker:1994ea}  & $9704\pm56$ & $|g_\tau/g_\mu|^2 = 0.993 \pm 0.006 $ \\ 
$\Gamma(\tau \rightarrow K  \nu)/\Gamma(K \rightarrow \mu \nu)$ & $480\pm 1$~\cite{Decker:1994ea} & $469 \pm 7$ & $|g_\tau/g_\mu|^2=0.977 \pm 0.015$ \\ 
$\Gamma(D_s \rightarrow \tau  \nu )/\Gamma(D_s \rightarrow \mu \nu)$ & 9.76~\cite{Agashe:2014kda} & $10.0\pm0.6$ & $|g_\tau/g_\mu|^2=1.02\pm0.06$ \\ 
$\Gamma(\bar{B}\rightarrow D^*\tau \nu)/\Gamma(\bar{B}\rightarrow D^* \mu \nu)$ & $0.252\pm 0.003$~\cite{Fajfer:2012vx} & $0.336\pm0.040$~\cite{Aaij:2015yra} & $|g_\tau/g_\mu|^2= 1.333\pm 0.159$ \\ 
\hline \hline
$\Gamma(\tau \rightarrow \pi  \nu)/\Gamma(\pi \rightarrow e \nu)$ & $(7.91 \pm 0.01)\times 10^7$~\cite{Decker:1994ea,Cirigliano:2007xi} & $(7.89 \pm 0.05)\times 10^7$ & $|g_\tau/g_e|^2 = 1.000\pm0.007$ \\
$\Gamma(\tau \rightarrow K  \nu)/\Gamma(K \rightarrow e \nu)$ & $(1.940\pm0.004)\times10^7$~\cite{Decker:1994ea,Cirigliano:2007xi} & $(1.89\pm0.03)\times10^7$  & $|g_\tau/g_e|^2 = 0.974 \pm 0.015$ \\ 
$\Gamma(W\rightarrow \tau \nu)/\Gamma(W\rightarrow e \nu )$ & 0.999~\cite{Kniehl:2000rb} & $1.063\pm0.027$ & $|g_\tau/g_e|^2 = 1.063 \pm 0.027$ \\ 
\hline  \hline
$\Gamma(\bar{B}\rightarrow D^*\tau \nu)/\Gamma(\bar{B}\rightarrow D^* \ell \nu)$ & $0.252\pm 0.003$~\cite{Fajfer:2012vx} & $0.318\pm0.024$~\cite{Huschle:2015rga,Lees:2012xj} & $2|g_\tau|^2/(|g_e|^2+|g_\mu|^2)= 1.262\pm 0.096$ \\ 
$\Gamma(\bar{B}\rightarrow D\tau \nu)/\Gamma(\bar{B}\rightarrow D \ell \nu)$ & $0.299\pm 0.011$~\cite{Lattice:2015rga} & $0.406\pm0.050$~\cite{Huschle:2015rga,Lees:2012xj} & $2|g_\tau|^2/(|g_e|^2+|g_\mu|^2)= 1.359\pm0.171$ \\ 
\hline
\end{tabular}
\caption{Tests for lepton universality that involve a neutrino.  All measurements are taken from, or estimated with information provided in, Ref.~\cite{Agashe:2014kda}, except for $\bar{B}$ decays, which are taken from Refs~\cite{Aaij:2015yra, Huschle:2015rga, Lees:2012xj}.   The values of $|g_\ell/g_{\ell'}|^2$ quoted are the factors by which the SM is multiplied to  match the experimental central value, where a $\chi^2$ function is used to estimate a 68.3\% CL error bar, combining both experimental and theoretical uncertainties.  In the SM, $|g_\mu/g_e|^2$, $|g_\tau/g_\mu|^2$ and $|g_\tau/g_e|^2$ are all predicted to be exactly unity. }
\label{leptuniv}
\end{table}
\end{center}

\begin{figure}[htbp]
\begin{center}
\includegraphics[width=0.47\linewidth]{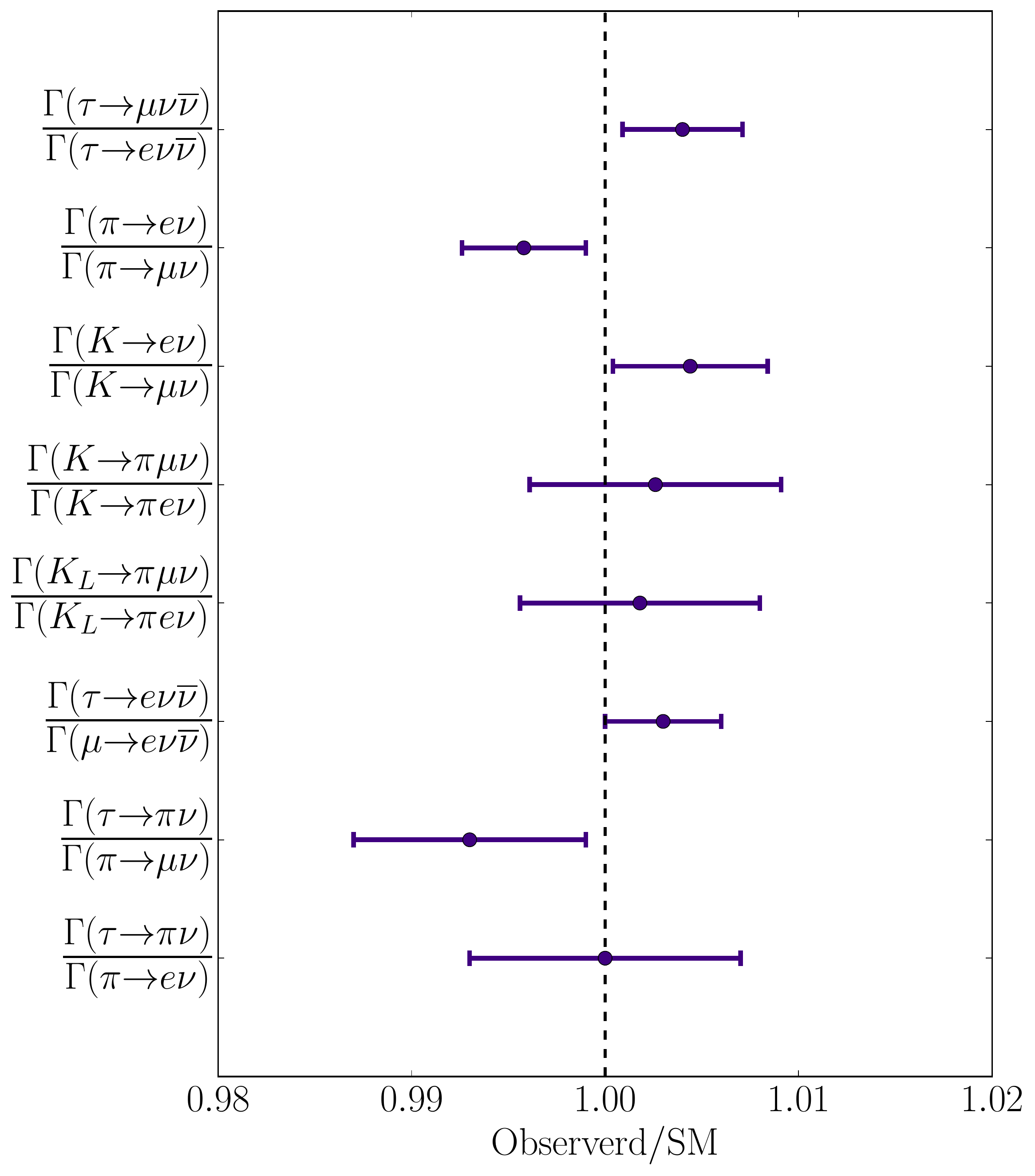} \hspace{0.2in}
\includegraphics[width=0.47\linewidth]{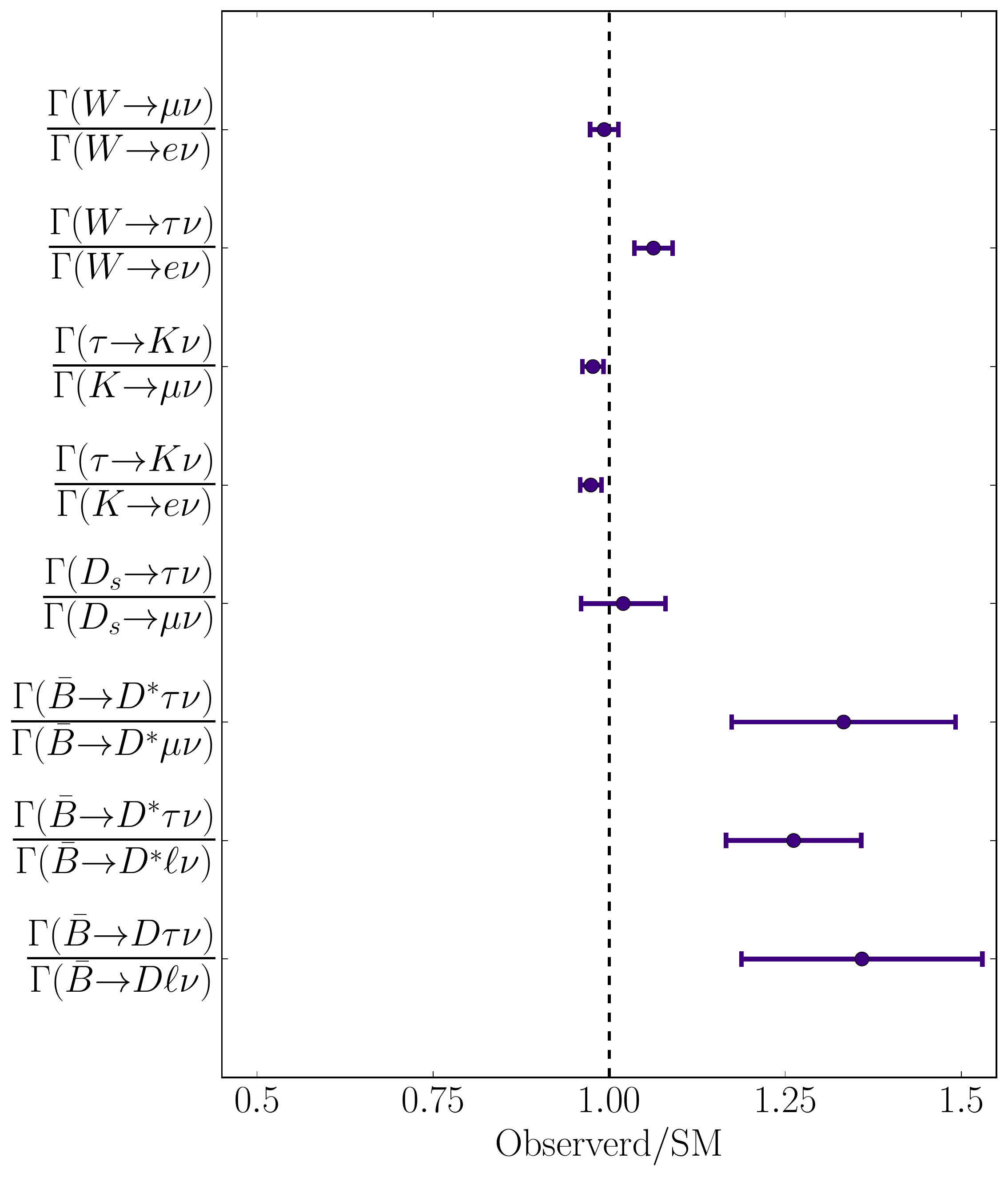}
\caption{ The ratio of the data and the SM expectation of the observables listed in Table~\ref{leptuniv}.  }
\label{leptunivplots}
\end{center}
\end{figure}

We quantify the differences between the experimental measurements and the SM predictions listed in Table~\ref{leptuniv} by multiplying the SM prediction by $|g_\ell/g_{\ell'}|^2$ and calculating the value it ought to have such that the central value of the theoretical prediction exactly matches the central value of the experimental measurement (the comparison between experiment and the SM expectation is also shown in Fig.~\ref{leptunivplots}).  Then, we use a $\chi^2$ function to estimate an error bar (68.3\% CL) on each individual value of $|g_\ell/g_{\ell'}|^2$, including both experimental and theoretical uncertainties.  We take  each $|g_\ell/g_{\ell'}|^2$ in Table~\ref{leptuniv} as being statistically independent from the others. 
If one combines all  values of $|g_\ell/g_{\ell'}|^2$ listed in Table~\ref{leptuniv}, the comparison to the SM expectation of $|g_\ell/g_{\ell'}|^2=1$ yields $p \simeq 4.6\times10^{-3}$ ($\chi^2/\text{dof}\simeq 34.5/16$), assuming each experimental value in Table~\ref{leptuniv} counts for only a single degree of freedom.  
The data are not consistent with lepton universality (at a little less than the $3\sigma$ level). The discrepancy between the data and the SM predictions could indicate that one or the other have underestimated systematic uncertainties, especially for the measurements or predictions of $\bar{B}$ decays and  hadronic $\tau$ decays.
 
Another possibility is that the existence of a heavy neutrino could be contributing to the observations, thus introducing a tension with the SM predictions.
If the heavy neutrino is light enough to be produced in a decay process, its presence will affect not only decay rates but also the content and kinematic distributions of particles in the final state. In order to maintain decay-model independence and circumvent nontrivial experimental considerations, we apply the individual lepton universality constraints only when the mass of the heavy neutrino is large enough that its production is kinematically forbidden.   In this case, the expressions for $|g_\mu/g_e|^2$, $|g_\tau/g_\mu|^2$, and $|g_\tau/g_e|^2$ in terms of the neutrino mixing-matrix elements can be found in Table~\ref{leptunivconst}. It should be noted that the model including a heavy neutrino is not identical to the hypothesis of lepton nonuniversality because the expressions for $|g_\ell/g_{\ell'}|^2$ in terms of the mixing-matrix element are not the same for every observable.
If the mass of the heavy neutrino is greater than the mass of the $W$ boson, and if $|U_{e4}|^2$, $|U_{\mu4}|^2$, and $|U_{\tau4}|^2$ are permitted to vary independently, then the best fit yields $p=2.9\times10^{-3}$ ($\chi^2_\text{min}/\text{dof}\simeq 31.4/13$), again, assuming each experimental value in Table~\ref{leptuniv} counts for a single degree of freedom. Nonzero $|U_{e4}|^2$, $|U_{\mu4}|^2$, and $|U_{\tau4}|^2$ do not provide a better fit to the data than the SM. 

We estimate the following marginalized limits on $|U_{e4}|^2$, $|U_{\mu4}|^2$, and $|U_{\tau4}|^2$ at 90\% CL, relative to $\chi^2_\text{min}$, when $m_4 > m_W$:  $|U_{e4}|^2< 5.9\times10^{-3}$, $|U_{\mu4}|^2< 2.5\times10^{-3}$, $|U_{\tau4}|^2< 5.9\times10^{-3}$, which depend on no assumptions about the heavy neutrino besides Eq.~(\ref{nulin}).  
These results can be distinguished from those already in the literature, because we use ratios of decay rates (which permit cancellations of systematic uncertainties),  marginalize over variables when quoting limits (instead of letting only one matrix element be nonzero at a time), use up-to-date measurements and predictions, include all available data (16 observables versus a few), and only quote limits when the value of $m_4$ is large enough that the decay into $\nu_4$ is kinematically forbidden.  This leads to bona fide model-independent limits on mixing-matrix elements (for other analyses see, for example, Refs.~\cite{Atre:2009rg, Bertoni:2014mva, delAguila:2008pw}). 
The full results, when 10 eV $\lesssim m_4\lesssim$ 1 TeV, are shown as the red-dashed line in Figs.~\ref{ue4},~\ref{umu4}, and~\ref{utau4}, which are the  99\% CL limits on $|U_{e4}|^2$, $|U_{\mu4}|^2$, and $|U_{\tau4}|^2$, respectively.  

\begin{table}[h]
\begin{tabular}{| l | c |}
\hline 
Observable & $|g_\ell/g_{\ell'}|^2$  \\ \hline \hline
$\Gamma(\tau\rightarrow \mu \nu \overline{\nu})/\Gamma(\tau\rightarrow e \nu \overline{\nu})$ & $|g_\mu/g_e|^2 = (1-|U_{\tau4}|^2 - |U_{\mu4}|^2)/(1-|U_{\tau4}|^2 - |U_{e4}|^2)$ \\ 
$\Gamma(\pi\rightarrow e \nu)/\Gamma(\pi\rightarrow \mu \nu )$ & $|g_e/g_\mu|^2 =(1-|U_{e4}|^2)/(1-|U_{\mu4}|^2)$  \\ 
$\Gamma(K\rightarrow e \nu)/\Gamma(K\rightarrow \mu \nu )$ & $|g_e/g_\mu|^2 =(1-|U_{e4}|^2)/(1-|U_{\mu4}|^2)$ \\ 
$\Gamma(K\rightarrow \pi \mu \nu)/\Gamma(K\rightarrow \pi e \nu )$ & $|g_\mu/g_e|^2 =(1-|U_{\mu4}|^2)/(1-|U_{e4}|^2)$ \\ 
$\Gamma(K_L\rightarrow \pi \mu \nu)/\Gamma(K_L\rightarrow \pi e \nu )$ & $|g_\mu/g_e|^2 =(1-|U_{\mu4}|^2)/(1-|U_{e4}|^2)$ \\ 
$\Gamma(W\rightarrow \mu \nu)/\Gamma(W\rightarrow  e \nu )$ & $|g_\mu/g_e|^2 =(1-|U_{\mu4}|^2)/(1-|U_{e4}|^2)$ \\ 
\hline \hline
$\Gamma(\tau \rightarrow e \nu \overline{\nu})/\Gamma(\mu \rightarrow e \nu \overline{\nu})$ & $|g_\tau/g_\mu|^2 = (1-|U_{\tau4}|^2-|U_{e4}|^2)/(1-|U_{\mu4}|^2-|U_{e4}|^2)$ \\
$\Gamma(\tau \rightarrow \pi  \nu)/\Gamma(\pi \rightarrow \mu \nu)$ & $|g_\tau/g_\mu|^2 = (1-|U_{\tau4}|^2)/(1-|U_{\mu4}|^2)$ \\
$\Gamma(\tau \rightarrow K  \nu)/\Gamma(K \rightarrow \mu \nu)$ & $|g_\tau/g_\mu|^2 = (1-|U_{\tau4}|^2)/(1-|U_{\mu4}|^2)$ \\
$\Gamma(D_s \rightarrow \tau  \nu)/\Gamma(D_s \rightarrow \mu \nu)$ & $|g_\tau/g_\mu|^2 = (1-|U_{\tau4}|^2)/(1-|U_{\mu4}|^2)$ \\
$\Gamma(\bar{B}\rightarrow D^*\tau\nu)/\Gamma(\bar{B}\rightarrow D^*\mu\nu)$ & $|g_\tau/g_\mu|^2 = (1-|U_{\tau4}|^2)/(1-|U_{\mu4}|^2) $ \\ 
\hline \hline
$\Gamma(\tau \rightarrow \pi  \nu)/\Gamma(\pi \rightarrow e \nu)$ & $|g_\tau/g_e|^2 = (1-|U_{\tau4}|^2)/(1-|U_{e4}|^2)$ \\
$\Gamma(\tau \rightarrow K  \nu)/\Gamma(K \rightarrow e \nu)$ & $|g_\tau/g_e|^2 = (1-|U_{\tau4}|^2)/(1-|U_{e4}|^2)$ \\
$\Gamma(W\rightarrow \tau \nu)/\Gamma(W\rightarrow e \nu )$ & $|g_\tau/g_e|^2 = (1-|U_{\tau4}|^2)/(1-|U_{e4}|^2)$ \\ 
\hline \hline
$\Gamma(\bar{B}\rightarrow D^{(*)}\tau\nu)/\Gamma(\bar{B}\rightarrow D^{(*)}\ell\nu)$ & $2|g_\tau|^2/(|g_e|^2+|g_\mu|^2) = 2(1-|U_{\tau4}|^2)/(2-|U_{e4}|^2-|U_{\mu4}|^2) $ \\ 
\hline
\end{tabular}
\caption{The first-order expressions (excluding $\mathcal{O}(|U_{\alpha4}|^4)$ and $\mathcal{O}(|U_{\alpha4}|^2|U_{\beta4}|^2)$ terms) for the values of $|g_\ell/g_{\ell'}|^2$ in Table~\ref{leptuniv} for each observable if the SM is augmented by a single heavy neutrino, assuming the heavy neutrino mass is large enough that the decay into it is kinematically forbidden.  These expressions do not consider the subsequent decays of a final-state tau.}
\label{leptunivconst}
\end{table}

\subsection{Invisible $Z$ Boson Decays ($m_4 > M_Z$)}
\label{zdecays} 

Precise measurements of the invisible $Z$ boson width can provide information on a heavy, mostly sterile neutrino.  A convenient way to parametrize these measurements is to define the quantity (as done in Ref.~\cite{ALEPH:2005ab})
\begin{equation}
N_\nu \equiv \frac{\Gamma(Z\rightarrow\text{inv})}{\Gamma(Z\rightarrow \ell \ell)}\Bigg|_\text{exp} \times \frac{\Gamma(Z\rightarrow\ell\ell)}{\Gamma(Z\rightarrow \nu\nu)}\Bigg|_\text{SM}, 
\end{equation}
where $\Gamma(Z\rightarrow\text{inv})/\Gamma(Z\rightarrow \ell \ell)$ is the measured ratio of the $Z$-boson decay rate to invisible particles and a given flavor of charged leptons ($\ell\ell = ee,$ $\mu\mu$, $\tau\tau$). The value of $\Gamma(Z\rightarrow\nu\nu)/\Gamma(Z\rightarrow \ell\ell)$ is predicted in the SM to be $1.9913\pm0.0008$~\cite{Agashe:2014kda}.   The SM expectation is $N_\nu=3$, and the LEP experiments measure $N_\nu=2.9840 \pm 0.0082$~\cite{ALEPH:2005ab}.  This presents, roughly, a $2 \sigma$ inconsistency between the data and the SM expectation.   

The discrepancy between the data and the SM can be completely accounted for if there is a heavy neutrino that interacts with the $Z$-boson only through mixing with the three active neutrinos.  A model-independent statement can be made when $m_4>M_Z$, where the heavy neutrino cannot be produced in $Z$ boson decays.  This removes the need for assumptions regarding how $\nu_4$ could decay.  If so, the expected value of $N_\nu$ is modified to
\begin{eqnarray}
\label{nunum}
N_\nu &=& 3\left(1-|U_{e4}|^2-|U_{\mu4}|^2-|U_{\tau4}|^2\right) + \mathcal{O}\left(|U_{\alpha4}|^4\right),
\end{eqnarray}
Comparing Eq.~(\ref{nunum}) with the experimental value implies that $|U_{e4}|^2+|U_{\mu4}|^2+|U_{\tau4}|^2 < 9.5\times10^{-3}$ at 90\% CL (or $< 1.2\times10^{-2}$ at 99\% CL).  Loop corrections do not greatly affect this result~\cite{Fernandez-Martinez:2015hxa}.  Such a limit is weaker than those estimated from tests for lepton universality.  In Section~\ref{zdecaysallmass}, we will discuss a more general version of Eq.~(\ref{nunum}) for all values of $m_4$.

\subsection{Lepton-Flavor-Violating Decays}
\label{LFV}

Loop-induced lepton-flavor-violating decays of charged leptons, e.g., radiative decays, three-body decays, semileptonic decays, and $\mu-e$ conversion, are predicted to be extremely rare in the SM, far beyond experimental reach, due to the smallness of the light neutrino masses.  Current experimental limits on the rates of these processes can be found in Table~\ref{LFVconstraints}. If there is a heavy neutrino with mass $m_4 \gtrsim 1$ MeV, the rates of these processes can be enhanced, perhaps to the point of being observable.  The following discusses some details regarding the theoretical predictions:
\begin{itemize}
\item {\it Charged-lepton radiative decays}.  At one loop, the branching ratio for $\ell\rightarrow \ell'\gamma$ is~(see Refs.~\cite{Alonso:2012ji, Forero:2011pc} and many references therein):
\begin{eqnarray}
Br(\ell\rightarrow \ell'\gamma) \simeq \frac{\alpha_W^3 s_W^2}{256 \pi^3} \frac{m_\ell^5}{M_W^4 \Gamma_\ell} \Bigg| U_{\ell'4}^* U_{\ell 4}G\left( \frac{m_4^2}{M_W^2}\right) \Bigg|^2,
\end{eqnarray}
where $\alpha_W \equiv g_W^2/4\pi$, $\Gamma_\ell$ is the total decay rate of $\ell$, and
\begin{equation}
G(x) \equiv  \frac{x(1-6x+3x^2+2x^3-6x^2\ln x)}{4(1-x)^4}.
\end{equation}

\item {\it Three-body charged-lepton decays}.  Including the effects of a heavy neutrino, we use the one-loop expressions in Ref.~\cite{Ilakovac:1994kj} to calculate the rates for lepton-flavor-violating three-body decays of charged leptons, i.e., $\mu^-\rightarrow e^-e^+e^-$ and $\tau^-\rightarrow e^-e^+e^-,e^- e^+ \mu^-,e^- \mu^+ e^-,e^- \mu^+ \mu^-,\mu^- e^+ \mu^-,\mu^- \mu^+ \mu^-$, as listed in Table~\ref{LFVconstraints}. We include diagrams with two heavy neutrinos in the loop,\footnote{For this reason, our results are slightly different from some of those in Ref.~\cite{Alonso:2012ji}.}  whose contributions can be significant since they scale like $m_4^2/M_W^2$ when $m_4$ is large. Furthermore, when these contributions are ignored, the $Br(\tau^-\rightarrow e^+\mu^-\mu^-)$ and $Br(\tau^-\rightarrow \mu^+e^-e^-)$ are zero at one loop.

\item {\it $\mu-e$ conversion}.  We use the expressions from Ref.~\cite{Alonso:2012ji} to estimate limits from searches for $\mu-e$ conversion in nuclei (including diagrams with two heavy neutrinos in the loop).  Table~\ref{LFVconstraints} lists the experimental constraints on the normalized rates for $\mu-e$ conversion $R^Z_{\mu\rightarrow e}$ on $Z=$~Ti, Au, S, and Pb.  

\item {\it Muonium decays and transitions}.  Muonium-antimuonium (MU-$\overline{\text{MU}}$) transitions and MU$\rightarrow e^+e^-$ decays can have a significant contribution due to Majorana and Dirac heavy neutrinos.  However, the current experimental constraints on these observables give very weak limits on the values of $|U_{e4}|$ and $|U_{\mu4}|$.  See Ref.~\cite{Abada:2015oba} for a detailed discussion regarding these observables. 

\end{itemize}

\begin{table}[h]
\begin{tabular}{| c | c |}
\hline 
Observable & Exp.~Limit ($90\%$ CL) \\ \hline \hline
$Br(\mu^-\rightarrow e^-\gamma)$ & $<5.7\times10^{-13}$ \\ \hline
$Br(\tau^-\rightarrow e^-\gamma)$ & $< 3.3\times10^{-8} $\\ \hline 
$Br(\tau^-\rightarrow \mu^-\gamma)$ & $< 4.4\times10^{-8} $\\ \hline \hline
$Br(\mu^-\rightarrow e^-e^+e^-)$ & $<1.0\times10^{-12}$ \\ \hline
$Br(\tau^-\rightarrow e^-e^+e^-)$ & $< 2.7\times10^{-8}$ \\ \hline
$Br(\tau^-\rightarrow \mu^-\mu^+\mu^-)$ & $<2.1\times10^{-8} $ \\ \hline
$Br(\tau^-\rightarrow e^-\mu^+\mu^-)$ & $< 2.7\times10^{-8}$ \\ \hline
$Br(\tau^-\rightarrow \mu^-e^+e^-)$ & $< 1.8\times10^{-8}$ \\ \hline
$Br(\tau^-\rightarrow e^+\mu^-\mu^-)$ & $< 1.7\times10^{-8}$ \\ \hline
$Br(\tau^-\rightarrow \mu^+e^-e^-)$ & $< 1.5\times10^{-8}$ \\ \hline \hline
$R^\text{Ti}_{\mu\rightarrow e}$ & $< 4.3\times10^{-12}$ \\ \hline
$R^\text{Au}_{\mu\rightarrow e}$ & $< 7\times10^{-13}$ \\ \hline
$R^\text{S}_{\mu\rightarrow e}$ & $< 7\times10^{-11}$ \\ \hline
$R^\text{Pb}_{\mu\rightarrow e}$ & $< 4.6\times10^{-11}$ \\ \hline
\end{tabular}
\caption{Experimental limits at 90\% CL on leptonic processes that violate lepton flavor (values taken from Ref.~\cite{Agashe:2014kda}). }
\label{LFVconstraints}
\end{table}

A heavy neutrino can also induce lepton-flavor-violating hadronic tau decays, $\tau \rightarrow \ell +\text{hadrons}$, where $\ell = \mu, e$. We do not, however, include the constraints from these searches, for the following reasons.  One concern is that, to our knowledge, precise computations of $\tau \rightarrow \ell +\text{hadrons}$ have not yet been performed. Furthermore, we expect the inclusion of such constraints not to quantitatively impact our final results.  For example, assuming the existence of a heavy neutrino, we estimate, very roughly, that the branching ratio $Br(\tau^- \rightarrow \ell^- \pi^0)\sim Br(\tau\rightarrow e^-\mu^-e^+)\times 16\pi^2 f_{\pi}^2/m_\tau^2$, taking into account the pion decay constant and different phase-space integration.  Because limits on the branching ratios of $\tau^- \rightarrow \ell^- \pi^0$ and $\tau\rightarrow e^-\mu^-e^+$ are both $\mathcal{O}(10^{-8})$, and $16\pi^2 f_\pi^2/m_\tau^2 \sim 0.8$, it is unlikely that lepton-flavor-violating hadronic tau decays provide significant additional information beyond the purely leptonic lepton-flavor-violating three-body tau decays that we have included. For a complete list of the experimental results of searches of this type, see Ref.~\cite{Agashe:2014kda}. 

\begin{figure}[htbp]
\begin{center}
\includegraphics[width=0.65\linewidth]{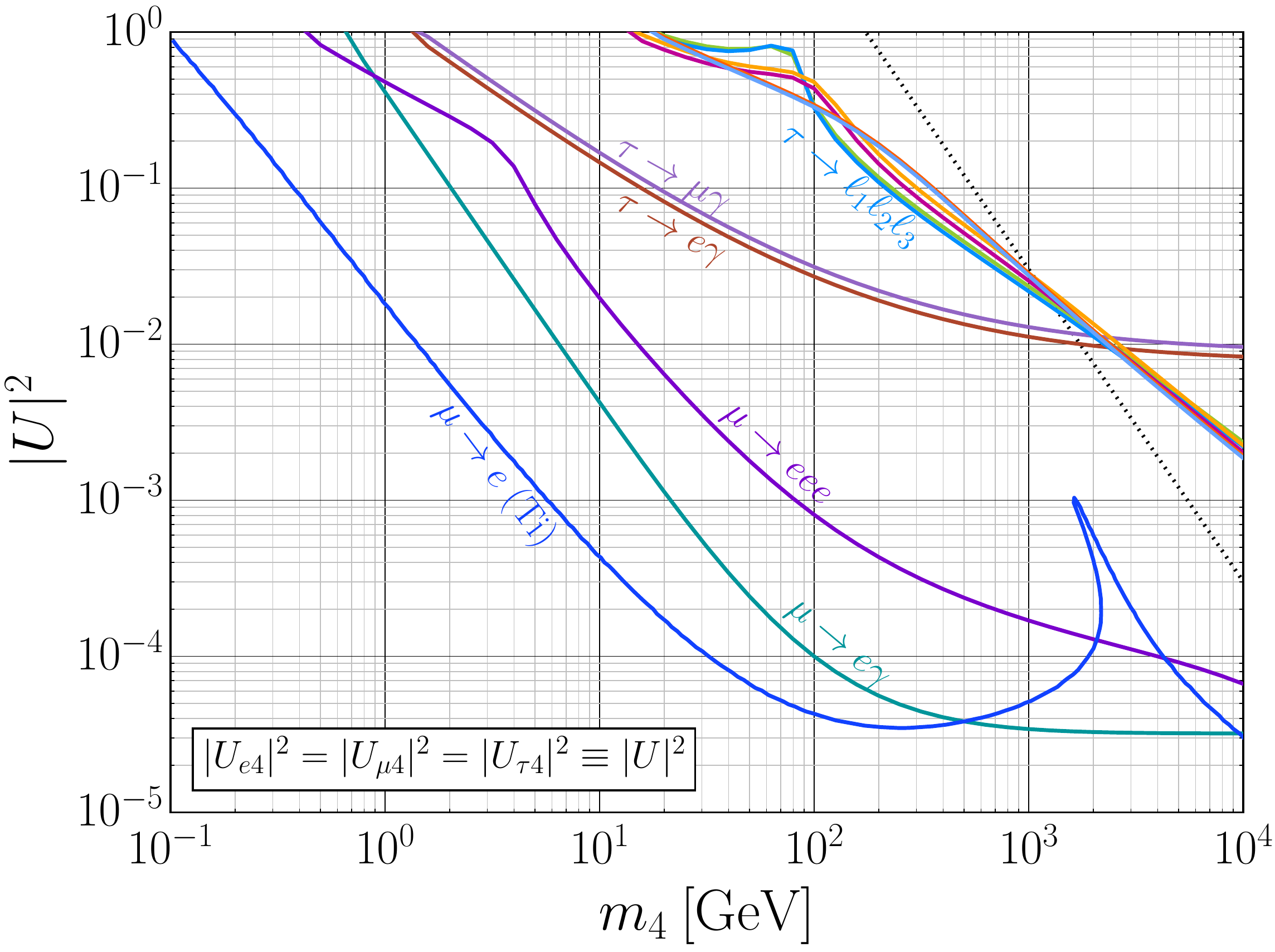}
\caption{ The 99\% CL limit on the neutrino mixing-matrix elements associated with a heavy neutrino, when $|U_{e4}|^2=|U_{\mu4}|^2=|U_{\tau4}|^2\equiv |U|^2$, using experimental constraints on radiative decays, three-body decays~\cite{Agashe:2014kda,Ilakovac:1994kj}, and $\mu-e$ conversion on Ti (which gives the strongest constraints of the $\mu-e$ various limits)~\cite{Agashe:2014kda,Alonso:2012ji}. The dashed black line corresponds to $|U|^2 = (174 \text{ GeV})^2/m^2_4$, to the right of which $|U|^2$ values are not expected to be theoretically accessible. See text for details.}
\label{LFVplot}
\end{center}
\end{figure}

We show in Fig.~\ref{LFVplot} the 99\% CL limits on the mixing-matrix elements associated with the heavy neutrino when $|U_{e4}|^2=|U_{\mu4}|^2=|U_{\tau4}|^2\equiv |U|^2$, utilizing a $\chi^2$ function and assuming it is zero when $|U|^2=0$.   
We also include a rough estimate for a theoretical upper bound on $|U|^2$: in certain Majorana neutrino models, there is a scaling between the mixing-matrix elements and the heavy neutrino mass, $U \sim y v/m_4$, where $y$ is a dimensionless coupling, $v \sim 174$ GeV is the vacuum expectation value of the SM Higgs boson, and $m_4$ is the mass of the heavy neutrino. If this is the case, $|U|^2\to 0$ as $m_4\to\infty$; $\nu_4$ ``decouples'' as its mass gets heavier, as expected. This naive upper bound on $|U|^2$ is shown as the dashed, black line in Fig.~\ref{LFVplot}, where we arbitrarily set the couplings $y$ to 1. If neutrinos are Dirac fermions, this estimate of the decoupling behavior need not apply. In general, the scale and behavior of $\nu_4$ decoupling depends on the details of the complete theory.

On their own, limits from charged-lepton-flavor violation can only constrain products of the matrix elements associated with heavy neutrino mixing and are, therefore, weaker than those estimated from lepton universality tests. In a nutshell, one can always satisfy the constraint on the product of two matrix elements by assuming one of them is very small.

\subsection{Neutrinoless Double Beta Decay}
\label{0nu2betadecay}

The observation of neutrinoless double beta decay ($0\nu2\beta$) would be a clear sign that neutrinos are Majorana fermions (though neither the converse nor inverse are true).  The conventional expression for the half life $T_{1/2}$ of $0\nu2\beta$ for a given nucleus is 
\begin{equation}
\frac{1}{T_{1/2}} = \frac{G_{0\nu} M_{0\nu}^2}{m_e^2} |m_{\beta\beta}|^2,
\end{equation}
where $G_{0\nu}$ is a phase-space factor that is $\sim\mathcal{O}(10^{14})$ yr$^{-1}$, and $M_{0\nu}$ is a nuclear matrix element.   Assuming one heavy neutrino exists, 
\begin{equation}
\label{mbetabeta}
|m_{\beta\beta}| = \Bigg| m_1 |U_{e1}|^2 e^{i\theta_1} + m_2 |U_{e2}|^2 e^{i\theta_2} + m_3 |U_{e3}|^2 e^{i\theta_3} + \left( \frac{m_4}{1-m_4^2/ p^2 }\right)  |U_{e4}|^2 e^{i\theta_4}  \Bigg|. 
\end{equation}
The phases $\theta_i$ represent linear combinations of phases present in the mixing matrix $U$, and $ p^2 $ is the virtuality of the neutrino exchanged in $0\nu2\beta$.  The phases are unknown and unconstrained, while the value of $p^2$ is also unknown, but it can be roughly estimated to be $p^2 \sim -(100-200 \text{ MeV})^2$~\cite{Mitra:2011qr}.  

\begin{figure}[htbp]
\begin{center}
\includegraphics[width=0.65\linewidth]{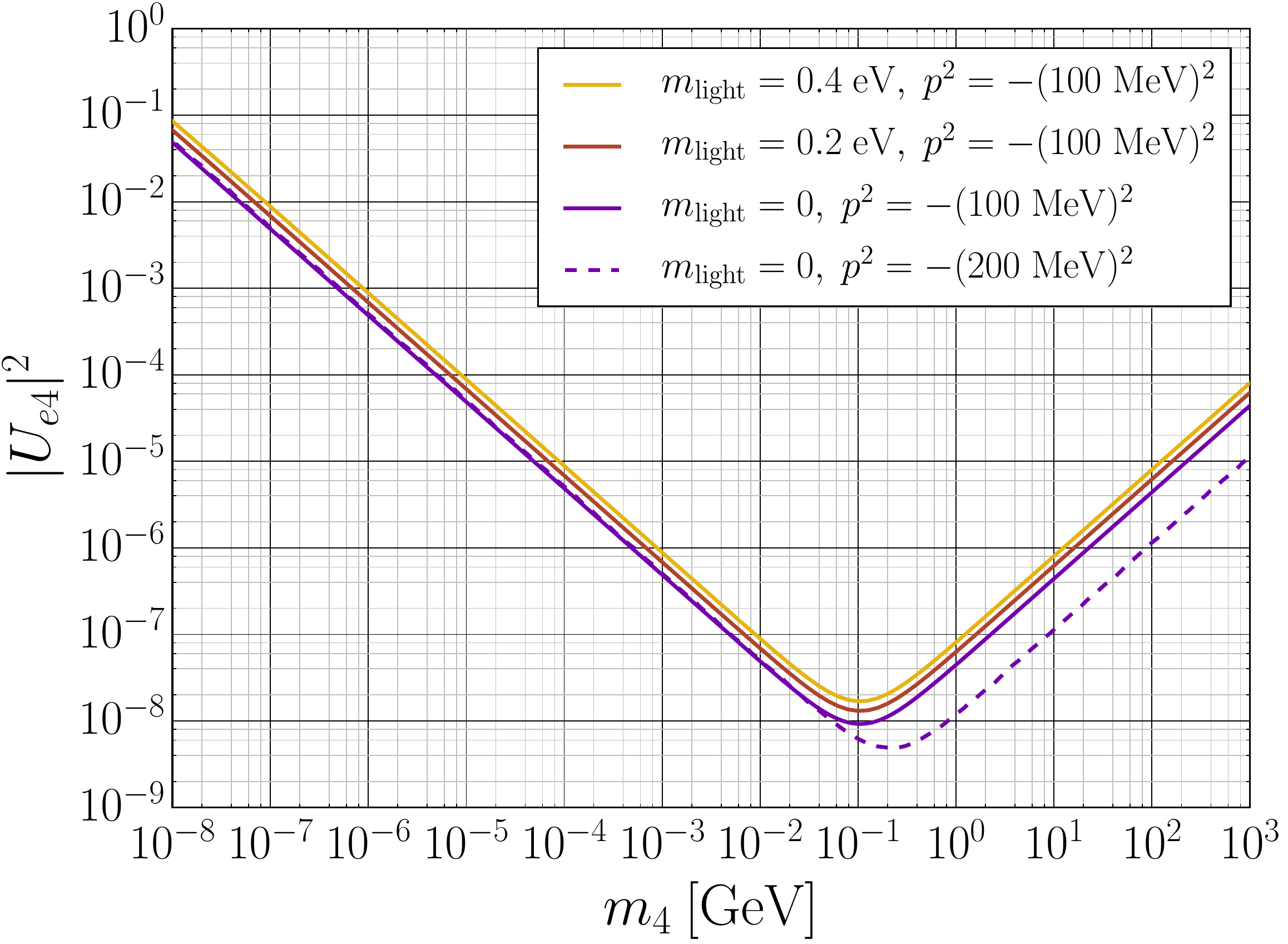}
\caption{ The 99\% CL upper limits on the value of $|U_{e4}|^2$ as a function of $m_4$ using the constraint $|m_{\beta\beta}| < 310$ meV~at 90\% CL \cite{Guzowski:2015saa} for different values of $m_\text{light}$ and $p^2$, using Eq.~(\ref{mbetabetared}).    }
\label{0nu2betaplot}
\end{center}
\end{figure}

The most conservative limit on a fourth neutrino is obtained when the phases and light masses are chosen to permit the strongest cancellation between terms in Eq.~(\ref{mbetabeta}), i.e., $\theta_1=\theta_2=\theta_3=0$, and $\theta_4=\pi$, while $m_1, m_2, m_3$ are as large as possible. The most stringent upper bound on the mostly active neutrino masses come from cosmological observables and are still consistent with the quasi-degenerate approximation, $m_1\approx m_2 \approx m_3 \equiv m_\text{light}$. The Planck collaboration reports that the sum of the relativistic neutrino masses is $\sum_i m_i < 0.23$ eV at 95\% CL~\cite{Ade:2015xua}, while $|m_3^2-m_1^2|$ and $m^2_2-m_1^2$ are known rather precisely and are both much smaller than $10^{-2}$~eV$^2$. Hence,
\begin{equation}
\label{mbetabetared}
|m_{\beta\beta}| > \Bigg| m_\text{light}(1-|U_{e4}|^2) - \left( \frac{m_4}{1-m_4^2/ p^2 }\right)  |U_{e4}|^2   \Bigg|.
\end{equation}
A combined analysis~\cite{Guzowski:2015saa} of the null results from searches for $0\nu2\beta$ by the GERDA~\cite{Agostini:2013mzu}, EXO-200~\cite{Albert:2014awa}, KamLAND-ZEN~\cite{Gando:2012zm}, CUORICINO~\cite{Andreotti:2010vj}, and NEMO-3~\cite{Arnold:2013dha} experiments places the limit $|m_{\beta\beta}| < 130-310$ meV at 90\% CL (the range is associated with the different estimations for the nuclear matrix elements).
If we consider the experimental constraint $m_{\beta\beta} < 310$ meV at 90\% CL and use a $\chi^2$ function to compare it with the expectation in Eq.~(\ref{mbetabetared}), assuming that $\chi^2=0$ when $|U_{e4}|^2=0$, then the 99\% CL limits on $|U_{e4}|^2$, as a function of $m_4$, can be found in Fig.~\ref{0nu2betaplot}, for different values of $m_\text{light}$ and $p^2$.  These limits depend on our assumption that the matrix elements $U_{\alpha i}$ are independent from one another and from the neutrino masses $m_i$. Very different limits are obtained under different circumstances. For example, in the Type-I see-saw model, bounds on $U_{e4}$ from $0\nu2\beta$ can be significantly weaker (for small $m_4$) or stronger (for large $m_4$) than the ones presented here (see, for example, \cite{deGouvea:2005er,deGouvea:2006gz,Blennow:2010th}).

In principle, there are other experimental lepton-number-violating constraints, e.g., lepton-number-violating $\mu^- - e^+$ conversion in nuclei and same-sign dilepton production at colliders.  Rare semileptonic meson decays, e.g. $K^+\to \pi^-\mu^+\mu^+$, can place model-independent limits on the existence of a heavy neutrino, analogous to neutrinoless double beta decay, but the current experimental limits are not yet strong enough to place meaningful constraints on $|U_{\alpha4}|$~\cite{Atre:2005eb}.

\subsection{Magnetic and Electric Dipole Moments}

The contributions from heavy neutrinos to the magnetic dipole moments of the charged leptons are beyond current experimental sensitivity.  New physics contributions to the magnetic moment of the electron are typically quite small, and the uncertainty associated with the magnetic moment of the muon is currently too large to meaningfully constrain the presence of a heavy neutrino (see, for example, Ref.~\cite{Abada:2014nwa}).  
Charged-lepton electric dipole moments can be induced from the presence of a heavy neutrino at two loops, but current experiments are not yet sensitive to these effects~\cite{Ng:1995cs,deGouvea:2005jj}.

\section{Constraints on an Invisible Heavy Neutrino}
\label{invconst}

In this section, we discuss constraints from experimental searches for a heavy neutrino, assuming that the heavy neutrino is produced as a final-state particle, and it either 1) does not decay on the length scale of the experiment, or 2) decays on the length scale of the experiment, but predominantly decays into invisible final-state particles.\footnote{These two criteria, phenomenologically speaking, result in the same interpretation of the data, except in the case of neutrino oscillations.}  

Before proceeding, it is useful to make some comments regarding expected $\nu_4$ lifetimes. Given that these interact, at least, via the weak interactions through their mixing with the active neutrinos, weak decays provide an upper bound to the lifetime of the heavy neutrino. For $m_4$ larger than the $W$-boson mass, $\nu_4$ decays are expected to be prompt. A rough estimate is $\tau_4< |U_{\alpha 4}|^{-2}\times 10^{-24}$~s for $m_4$ of order the top quark mass. Even for  $|U_{\alpha 4}|^2$ values smaller than any of the values accessible to the experiments discussed here (say, $|U_{\alpha 4}|^2\lesssim 10^{-10}$), $\nu_4$ is significantly shorter-lived than, e.g., $D$ mesons. For $m_4$ values in the GeV range or lower, $\nu_4$ decays like the tau or muon. Reasonable estimates are $\tau_4 < |U_{\alpha 4}|^{-2}\times 10^{-13}$~s for $m_4\sim m_{\tau}$, and $\tau_4 < |U_{\alpha 4}|^{-2}\times 10^{-6}$~s for $m_4\sim m_{\mu}$, keeping in mind that, in this mass range, the lifetime is proportional to $(m_4)^{-5}$. If $m_4$ is small enough that $\nu_4\to e^+e^-\nu$ is kinematically forbidden, the upper bound on the heavy neutrino lifetime is significantly higher.

A very rough rule of thumb is that if $m_4$ is smaller than a few hundred MeV and there are no new interactions, $\nu_4$  is stable at terrestrial experiments, and if $m_4$ is larger than 100 GeV, $\nu_4$ decays are prompt. In between, in the absence of new interactions, whether or not the heavy neutrinos decay within the length scale of a given experiment depends strongly on $m_4$ and $|U_{\alpha 4}|^2$. Our assumption that there are new interactions that lead the $\nu_4$ to decay predominantly invisibly sidesteps all issues associated with how and how quickly the heavy neutrinos decay. As discussed earlier, we view the constraints discussed below as most conservative.

\subsection{Kinematic Constraints}
\label{kinconst}

One can use the energy spectra of visible final-state particles in beta-decay, pion decay, kaon decay, muon decay, etc., to search for an invisible massive particle  in the final state~\cite{Shrock:1980ct, Shrock:1980vy}.  In Section~\ref{direct}, we itemize constraints on $|U_{e4}|^2$ and $|U_{\mu4}|^2$, taken directly from experiments. In Section~\ref{michel}, we outline our estimate for constraints on $|U_{e4}|^2 + |U_{\mu4}|^2$ from the precise measurements of the Michel electron energy spectrum in muon decay. 

\subsubsection{Direct Experimental Constraints}
\label{direct}

There are several dedicated kinematic searches that offer some of the best direct experimental constraints on an invisible heavy neutrino:

\begin{itemize}
\item Searches for heavy neutrinos via the kinematics of $\beta$ decay have been performed with $^{187}$Re~\cite{Galeazzi:2001py}, $^{3}$H~\cite{Hiddemann:1995ce, Belesev:2013cba}, $^{63}$Ni~\cite{Holzschuh:1999vy}, $^{35}$S~\cite{Holzschuh:2000nj}, $^{45}$Ca~\cite{Derbin:1997ut}, $^{64}$Cu~\cite{Schreckenbach:1983cg}, $^{20}$F (along with superallowed Fermi decays)~\cite{Deutsch:1990ut}.  The results of these searches are shown in Fig.~\ref{ue4} and exclude $|U_{e4}|^2 < \mathcal{O}(10^{-3})$ when $1 \text{ keV} \lesssim m_4 \lesssim 450$ keV.\footnote{The KATRIN experiment will be able to place very strong constraints on $|U_{e4}|^2$ using only the kinematics of $^3$H decay~\cite{deGouvea:2006gz,Formaggio:2011jg,Esmaili:2012vg,Mertens:2014nha, Mertens:2014osa}.}

\item An experiment performed at TRIUMF used  the kinematics of $\Gamma(\pi\rightarrow e\nu)$ to place limits on $|U_{e4}|^2<  \mathcal{O}(10^{-8})$ at 90\% CL for 10 MeV $\lesssim m_4 \lesssim$ 55 MeV~\cite{Britton:1992xv, Britton:1992pg}.  

\item The Brookhaven E949 experiment places the limit  $|U_{\mu4}|^2<\mathcal{O}(10^{-8})$ at 90\% CL by analyzing the kinematics of $K\rightarrow \mu \nu_4$ for 175 MeV $\lesssim m_4 \lesssim$ 300 MeV~\cite{Artamonov:2014urb}.  

\item The KEK E104 experiment constrains $|U_{e4}|^2<\mathcal{O}(10^{-6})$ at 90\% CL, when 135 MeV $\lesssim m_4 \lesssim$ 350 MeV, using the kinematics of $K\rightarrow e\nu_4$ decays~\cite{Yamazaki:1984sj}.

\item An experiment at KEK used the kinematics of $K\rightarrow \mu \nu$ decays to place the limit $|U_{\mu4}|^2 <\mathcal{O}(10^{-5})$ at  90\% CL for $70 \lesssim m_4 \lesssim 300$ MeV~\cite{Hayano:1982wu}.  

\item The authors of Ref.~\cite{Abela:1981nf} used pion decay to place limits on the ratio $|U_{\mu4}|^2/(1-|U_{\mu4}|^2)\equiv \Gamma(\pi \rightarrow \mu \nu_4) / \Gamma(\pi \rightarrow \mu \nu_i) \approx |U_{\mu4}|^2 < \mathcal{O}(10^{-4})$ for $10 \lesssim m_4 \lesssim 30$ MeV at 90\% CL. 

\end{itemize}
Absent from the above list are direct limits on $|U_{\tau4}|^2$.  Precision, high-statistics measurements of the kinematics of $\tau\rightarrow \nu + 3\pi$ are sensitive to nonzero $|U_{\tau4}|^2$ values when $100\text{ MeV}\lesssim m_4 \lesssim 1.2$ GeV, providing one of the only kinematic tools for placing limits on $|U_{\tau4}|^2$~\cite{Kobach:2014hea}.  To our knowledge, this analysis has not yet been performed.

\subsubsection{Michel Spectrum from Muon Decay}
\label{michel}

If there is a heavy neutrino in the final state of muon decay, then the energy spectrum of the final-state electron will change~\cite{Dixit:1983yz, Kalyniak:1981ye, Gninenko:2010pr, Shrock:1981wq}.  The differential muon decay rate is~\cite{Dixit:1983yz, Kalyniak:1981ye, Shrock:1981wq}
\begin{equation}
\label{michelspec}
\frac{d\Gamma(\mu\rightarrow e\overline{\nu} \nu)}{dx} = \frac{G_F^2 m_\mu^5}{192\pi^3}\mathcal{F}\left(x,\delta,\rho, |U_{e4}|^2, |U_{\mu4}|^2 \right) + \text{radiative corrections},
\end{equation}
where $x \equiv 2E_e/m_\mu$, $\delta \equiv m_4/m_\mu$, and 
\begin{equation}
\mathcal{F}\left(x,\delta,\rho, |U_{e4}|^2, |U_{\mu4}|^2 \right)\equiv \left(1-|U_{e4}|^2-|U_{\mu4}|^2\right)f(x, 0,\rho) + \left(|U_{e4}|^2+|U_{\mu4}|^2\right) f(x,\delta,\rho) ,
\end{equation}
\begin{equation} 
\label{fxdeltarho}
f(x,\delta,\rho) \equiv \frac{x^2}{2}\left[ 6(1-x) + \frac{4}{3}\rho (4x-3) - 3\delta^2 - \frac{3\delta^4 }{(1-x)^2}  - \frac{(x-3)\delta^6 }{(1-x)^3} \right] \Theta(1-x-\delta^2).
\end{equation}
The value of $\rho$ is predicted to be $\rho_\text{SM} = 3/4$ in the SM~\cite{Agashe:2014kda}, and the TWIST experiment measures it to be $\rho_\text{exp} = 0.74997 \pm 0.00026$~\cite{Bayes:2013esa}. 

To our knowledge, no experiment has fit the kinematic distributions of Michel electrons in muon decay to a model that includes a fourth neutrino.  In the absence of such a direct experimental result, we attempt to make a rough estimation of the limits on $|U_{e4}|^2+ |U_{\mu4}|^2$, considering that nonzero values of $|U_{\mu4}|^2$, and $|U_{e4}|^2$ could affect the fit to the data that determines the value of $\rho$.  
%
%
%
%
To do so, we define a $\chi^2$ function to compare the Michel electron energy spectrum to two functions, one where $\rho = \rho_\text{exp}$ and $\delta=0$, and another where $\rho = \rho_\text{SM}$, and $\delta$ is set to a given value.  We organize these distributions into electron energy bins with a width of 1~MeV (similar to the energy bins at TWIST~\cite{Grossheim:2009aa}).  The uncertainty in the denominator in the $\chi^2$ of each bin is the propagating uncertainty associated with $\rho_\text{exp}$.  For a given value of $\delta$, we vary the value $|U_{\mu4}|^2+|U_{e4}|^2$ in order to estimate limits.  We find that $|U_{\mu4}|^2+|U_{e4}|^2 < \mathcal{O}(10^{-3})$ for 10 MeV $\lesssim m_4 \lesssim$ 70 MeV at 99\% CL, as shown in Figs.~\ref{ue4} and~\ref{umu4}.  This result is similar in spirit to that found in Ref.~\cite{Gninenko:2010pr}, but we estimate limits for the full range of $m_4$ and note that the limits apply for the sum $|U_{\mu4}|^2+|U_{e4}|^2$ (as discussed in Ref.~\cite{Dixit:1983yz}), not just $|U_{\mu4}|^2$ alone.

\subsection{Invisible $Z$-Boson Decays}
\label{zdecaysallmass}

As first discussed in Section~\ref{zdecays}, the presence of a heavy neutrino can affect the measurement of the invisible width of the $Z$ boson.  We can easily extend the limits estimated Section~\ref{zdecays} for all values of $m_4$ with the assumption that if the heavy neutrino is produced in the decay of a $Z$ boson, then it is invisible to detection.  If so, the expression in Eq.~(\ref{nunum}) can be amended to 
\begin{eqnarray}
N_\nu  &\simeq& 3\left(1-|U_{e4}|^2-|U_{\mu4}|^2-|U_{\tau4}|^2\right)  \nonumber \\ 
&& +~ 3\left(|U_{e4}|^2+|U_{\mu4}|^2+|U_{\tau4}|^2\right) \left( 1- \frac{m_4^2}{M_Z^2} \right)^2 \left( 1+  \frac{1}{2}\frac{m_4^2}{M_Z^2} \right) \Theta(M_Z-m_4), 
\end{eqnarray}
up to order $|U_{\alpha4}|^4$.  The LEP experiments measure $N_\nu=2.9840 \pm 0.0082$~\cite{ALEPH:2005ab}, and the SM expectation is $N_\nu=3$.  The limits on $|U_{e4}|^2$, $|U_{\mu4}|^2$, and $|U_{\tau4}|^2$ are shown in Figs.~\ref{ue4},~\ref{umu4}, and~\ref{utau4}, respectively.

\subsection{Neutrino Oscillations}
\label{nuosc}

Neutrino oscillations can provide insight regarding the presence of a heavy neutrino.  
Because we focus on an fourth neutrino with mass $m_4 \gtrsim 10$ eV, the associated oscillations are typically too rapid to resolve experimentally.\footnote{As an aside, data from a handful of short-baseline neutrino oscillation experiments~\cite{Aguilar:2001ty,AguilarArevalo:2008rc,Mention:2011rk,Frekers:2011zz,Aguilar-Arevalo:2013pmq} disagree with our current understanding of neutrinos and can be interpreted as evidence for a fourth neutrino with mass $m_4\sim1$~eV.  A global analysis in Ref.~\cite{Kopp:2013vaa} reports best-fit values $\Delta m_{14}^2\approx 1\text{ eV}^2$, $|U_{e4}|^2\approx 0.02$, and $|U_{\mu4}|^2 \approx 0.03$.  However, these data are not entirely consistent with one another under the four-neutrino hypothesis~\cite{Kopp:2013vaa}, and the best-fit values are in disagreement with data from KARMEN~\cite{Armbruster:2002mp} and the combination of disappearance data from the MINOS and Bugey experiments~\cite{Timmons:2015lga}.  Proposed long- and short-baseline experiments, e.g., DUNE~\cite{Adams:2013qkq} or $\nu$STORM~\cite{Adey:2014rfv}, may be able to offer additional information regarding the possible existence of an additional light neutrino~\cite{Berryman:2015nua, deGouvea:2014aoa}.}  Even so, the oscillations of the three light neutrinos would be ``nonunitary," meaning the oscillation probability for only three light neutrinos is distinctly different from the oscillation probability for three light neutrinos and one heavy neutrino~\cite{Escrihuela:2015wra, Berryman:2015nua}. If $\nu_4$ has a negligible probability of decaying along its length of flight (and if the light neutrinos do not decay), then the probability for the oscillation $\nu_\alpha \rightarrow \nu_\beta$ is
%
\begin{equation}
\label{oscprobdecoup}
P_{\nu_\alpha\rightarrow\nu_\beta} \simeq \Big| \delta_{\alpha\beta}  - U_{\alpha 4}U_{\beta 4}^* + U_{\alpha 2}U_{\beta 2}^* \left( e^{-i \Delta_{12}}-1 \right) + U_{\alpha 3}U_{\beta 3}^* \left( e^{-i \Delta_{13}}-1 \right) \Big |^2 + |U_{\alpha 4}|^2|U_{\beta 4}|^2 .
\end{equation}
Here, $\Delta_{ij} \equiv 2.54 ( \Delta m^2_{ij}/1 \text{ eV}) \left( L/1 \text{ km} \right) \left( 1 \text{ GeV}/E_\nu \right)$, $\Delta m_{ij}^2\equiv m_j^2-m_i^2$, $L$ is the the experimental baseline, and $E_\nu$ is the beam energy.   The probability  $P_{\overline{\nu}_\alpha \rightarrow \overline{\nu}_\beta}$ is the same as in Eq.~(\ref{oscprobdecoup}), but the matrix elements would be complex conjugated.  

If $m_4 \gtrsim 10$ eV, and $\nu_4$ does not decay along the length of the oscillation experiment, then the KARMEN experiment constrains $4|U_{e4}|^2|U_{\mu4}|^2 < 1.3\times10^{-3}$ at 90\% CL~\cite{Armbruster:2002mp}, and the FNAL-E531 experiment constrains $4|U_{\mu4}|^2|U_{\tau4}|^2 < 4\times10^{-3}$ and $4|U_{e4}|^2|U_{\tau4}|^2 \lesssim 0.2$ at 90\% CL~\cite{Ushida:1986zn}.  Because both KARMEN and FNAL-E531 utilize pion beams, these constraints hold up to $m_4 \sim 1$ MeV, beyond which the phase-space suppression associated with the production of a heavy neutrino begins to take effect. These are some of the only constraints for $10\text{ eV} \lesssim m_4 \lesssim 1 \text{ MeV}$. 
If instead the heavy neutrino had some probability to decay along its flight path, then 
these constraints do not apply, and the constraints would have to be experimentally recalculated. 

Similar to searches for charged-lepton-flavor violation, the experiments above constrain only products of $|U_{\alpha 4}|^2|U_{\beta 4}|^2$ and hence do not lead to strong constraints on individual $|U_{\alpha 4}|^2$. Indeed, for small enough $m_4$, when only these constraints are applicable, the neutrino oscillation data discussed in this subsection cannot rule out the possibility that $|U_{\mu 4}|^2$ or $|U_{\tau 4}|^2$ are one. We return to this issue in Section~\ref{combo}.


\subsection{Cosmology}

If a heavy neutrino is in thermal equilibrium in the early universe, it can have an effect on cosmological observables, e.g., the Hubble constant, the primordial abundance of light nuclei, the cosmic microwave background (CMB), supernova luminosities, baryon acoustic oscillations (BAOs), and the large-scale distribution of galaxies.  Because we do not utilize a full model of sterile-neutrino interactions, an in-depth analysis of cosmological constraints is beyond the scope of our present analysis.  However, we do comment that if $\nu_4$ decays on a time scale sufficiently before big-bang nucleosynthesis (BBN) ($t_\text{BBN}\sim 0.1$ s), then very strong constraints from big-bang nucleosynthesis, CMB, BAOs, etc., can be significantly weakened, if not removed altogether (see, for example, Refs.~\cite{Smirnov:2006bu, Vincent:2014rja, Drewes:2015iva}). We qualitatively comment on bounds from cosmology in Section~\ref{combo}.

\section{Global Combination and Discussion}
\label{combo}

\begin{figure}[htbp]
\centering
\subfigure[]{\label{ue4}\includegraphics[width=.95\linewidth]{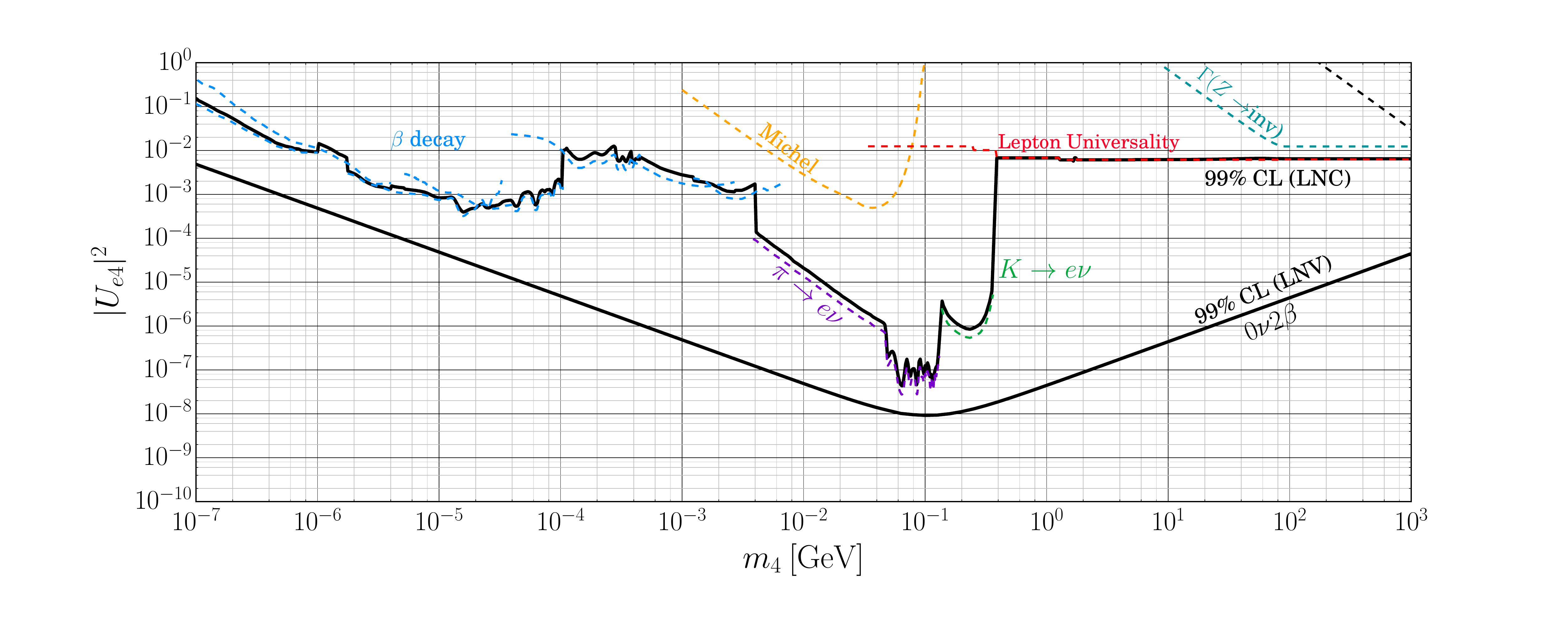}}
\subfigure[]{\label{umu4}\includegraphics[width=.95\linewidth]{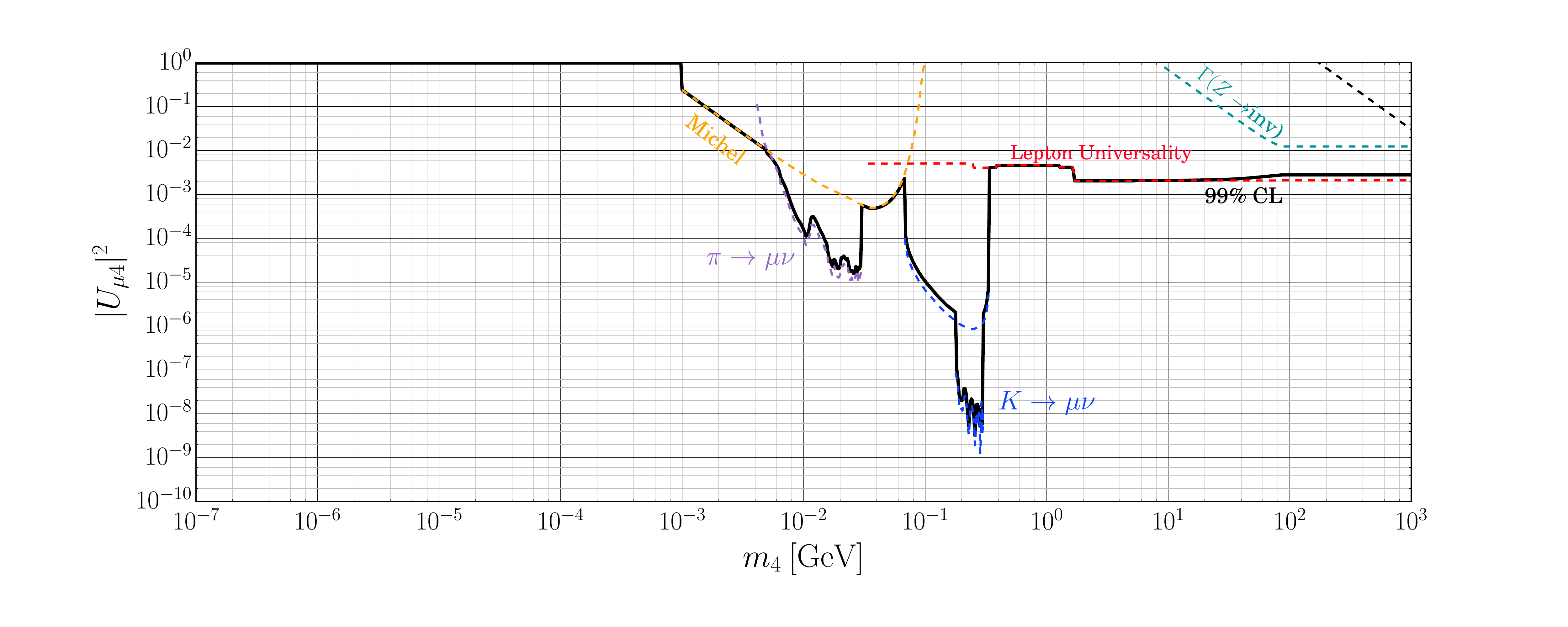}}
\subfigure[]{\label{utau4}\includegraphics[width=.95\linewidth]{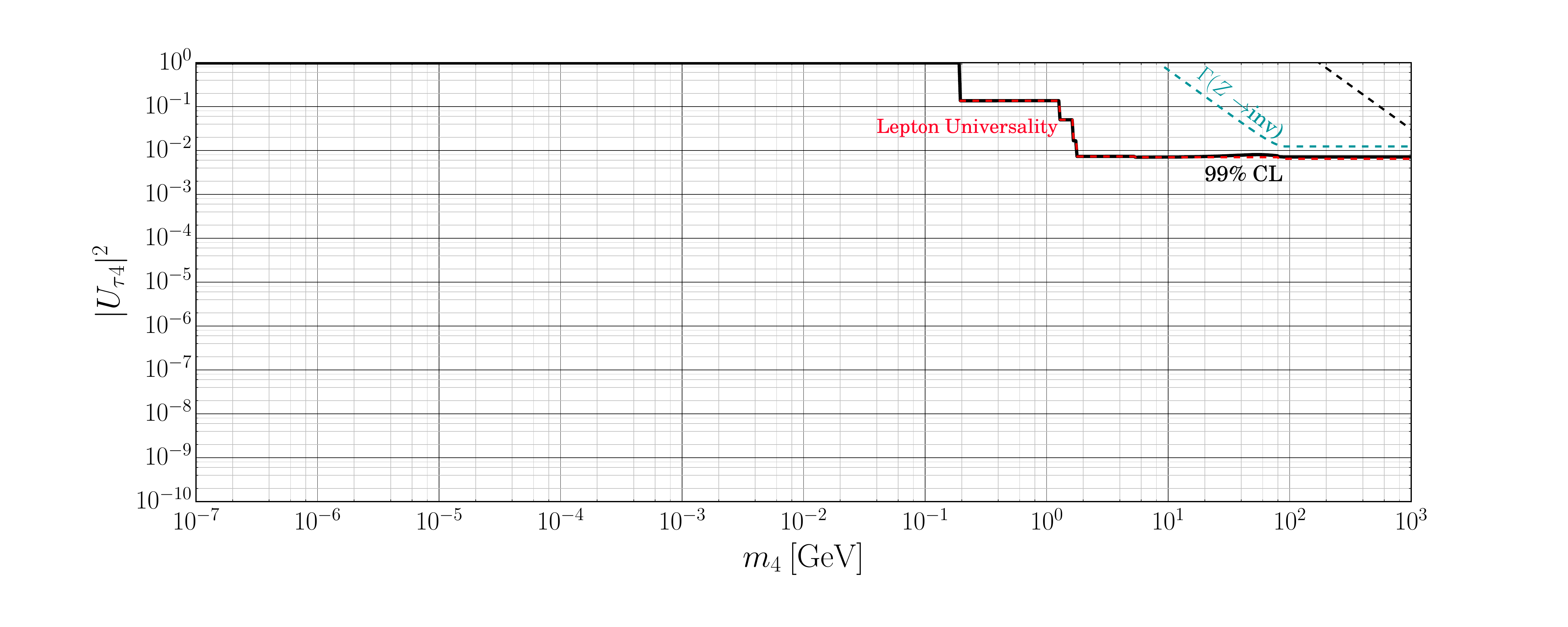}}
\caption{ The global 99\% CL upper limits on the value of (a) $|U_{e4}|^2$, (b) $|U_{\mu4}|^2$, and (c) $|U_{\tau4}|^2$ as a function of $m_4$.  See text for details and sources of each constraint.  The black dashed line corresponds to $|U|^2 =v^2/m_4^2$, to the right of which $|U|^2$ values are not expected to be theoretically accessible. See text for details.}
\label{diagonallimits}
\end{figure}

In order to combine all the constraints in Sections~\ref{modelindepconstraints} and~\ref{invconst}, we choose to define a $\chi^2$ function for each observable and for a given value of $m_4$ in the range $10\text{ eV} \lesssim m_4 \lesssim 1$ TeV.  In so doing, we make the following choices/assumptions: 
\begin{itemize}

\item In order to estimate {\it conservative} results, we make a phenomenological assumption that the heavy neutrino is invisible to detection, i.e., it either is long-lived relative to the scale of the experiment, or it decays quickly to other light species.

\item We apply the constraints from lepton universality tests (Section~\ref{lepunitests}) when $m_4$ is too large to be produced in the decay of the parent particle. This is done in order to avoid the effects of producing a massive neutrino, which can affect experimental measurements in a nontrivial way, e.g., reducing the momentum of visible particles to the point where they no longer pass event selection criteria.   

\item The constraints on $N_\nu$ from invisible $Z$-boson decays (Sections~\ref{zdecays} and~\ref{zdecaysallmass}) are applied for all values of $m_4$.

\item In order to quote a conservative bound, we choose $p^2=-(100\text{ MeV})^2$, $m_\text{light}=0.05$ eV, and $|m_{\beta\beta}|<310$ meV (90\% CL) when applying the constraints from neutrinoless double beta decay (Section~\ref{0nu2betadecay}).  Furthermore, we choose the associated $\chi^2$ to be zero when $|U_{e4}|^2=0$.

\item The constraints from $\mu\rightarrow e\gamma$, $\tau\rightarrow \ell\gamma$, $\mu\rightarrow3e$, $\tau\rightarrow \ell_1\ell_2\ell_3$, $\mu-e$ conversion on Ti (Section~\ref{LFV}), the limits from kinematic searches (Section~\ref{kinconst}), muon decay spectrum (Section~\ref{michel}), and neutrino oscillations (Section~\ref{nuosc}) are utilized assuming the individual $\chi^2$ functions are zero when all $|U_{\alpha i}|^2=0$.  
\end{itemize}
In Figs.~\ref{ue4},~\ref{umu4},~\ref{utau4}, the marginalized 99\% CL limits on $|U_{e4}|^2$, $|U_{\mu4}|^2$, and $|U_{\tau4}|^2$, respectively, are shown as a solid black line.  The reason why these global limits do not follow perfectly the individual limits (shown as dashed colored lines) in all places is that we have a consistent 99\% CL limit in our global combination, while limits from experiments are often quoted at 90\% and 95\% CL.
There is no difference between the limits on $|U_{\mu4}|^2$ and $|U_{\tau4}|^2$ when lepton-number violation is permitted, i.e., the constraint from neutrinoless double beta decay is included.  However, the choice of including limits from $0\nu2\beta$-decay has a very strong effect on the limits associated with $|U_{e4}|^2$, which can be seen in Fig.~\ref{ue4}.

\begin{figure}[htbp]
\centering
\subfigure[]{\label{ue4umu4100}\includegraphics[width=0.4\textwidth]{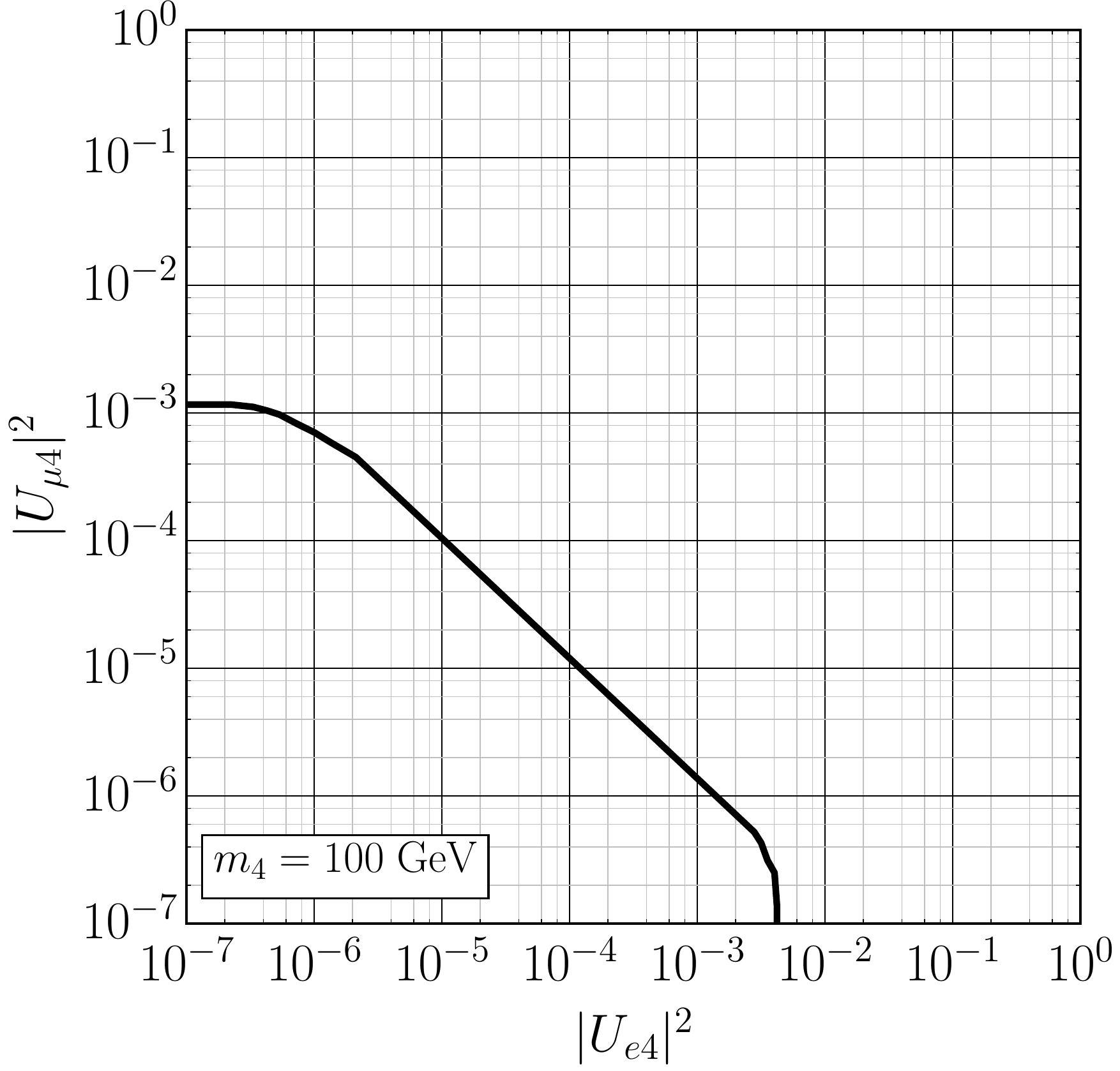}}
\subfigure[]{\label{ue4utau4100}\includegraphics[width=0.4\textwidth]{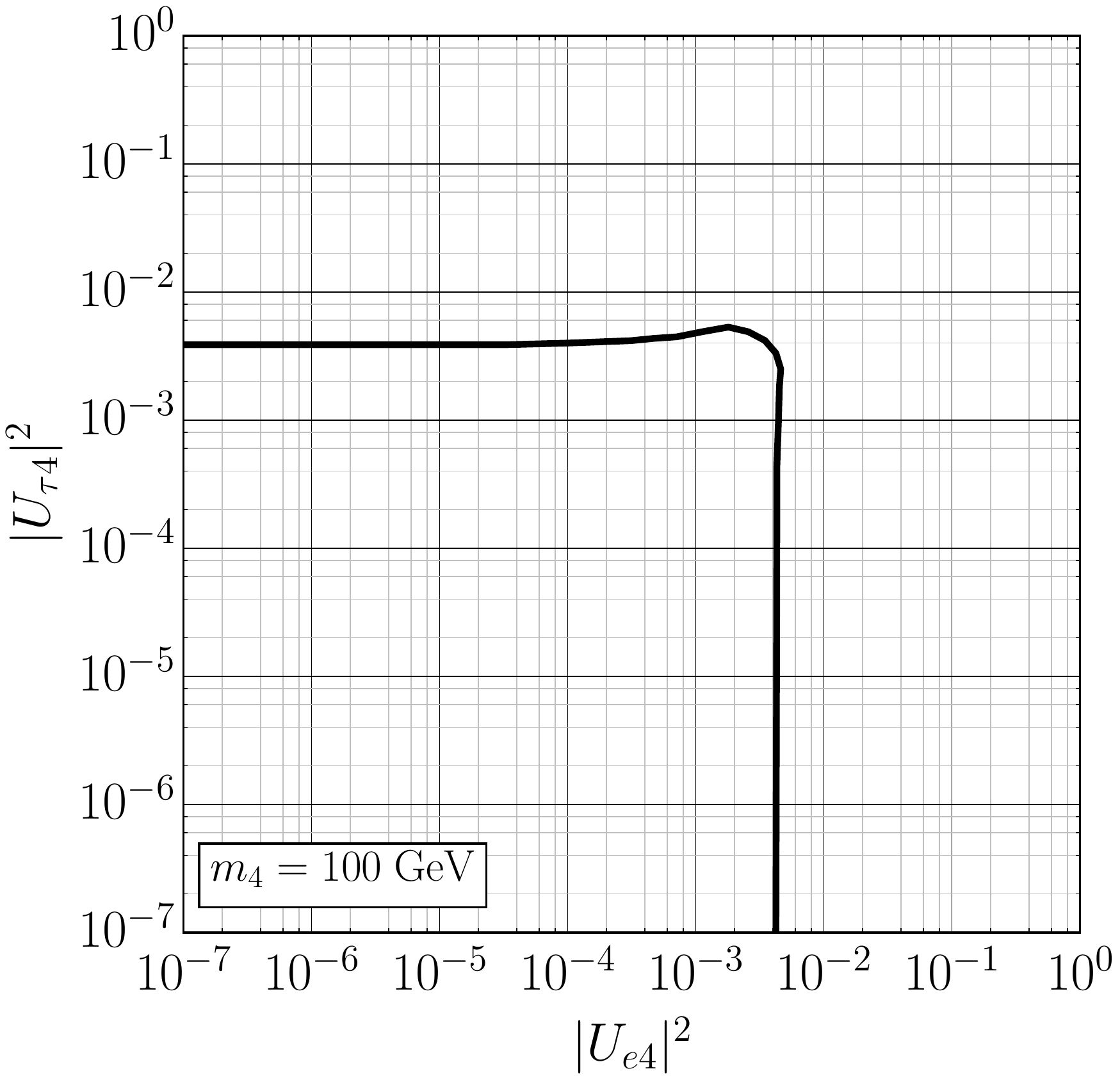}}\\
\subfigure[]{\label{umu4tau4100}\includegraphics[width=0.4\textwidth]{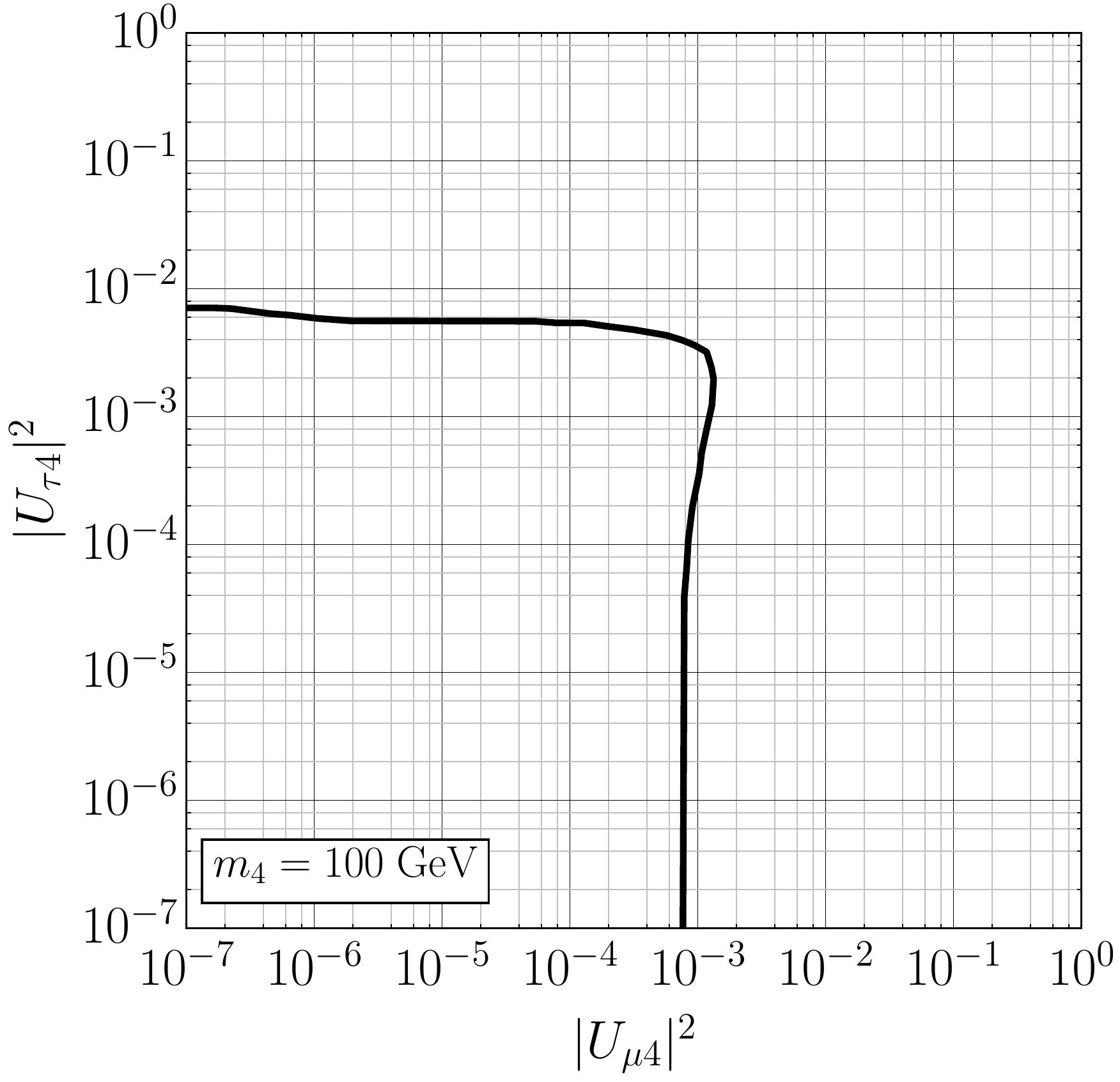}}
\caption{ The two-dimensional 99\% CL upper limits on (a) $|U_{e4}|^2$ vs.~$|U_{\mu4}|^2$, (b) $|U_{e4}|^2$ vs.~$|U_{\tau4}|^2$, and (c) $|U_{\mu4}|^2$ vs.~$|U_{\tau4}|^2$, when $m_4 = 100$ GeV, assuming lepton number conservation.  }
\label{offdiagonal_100GeV}
\end{figure}

\begin{figure}[htbp]
\centering
\subfigure[]{\label{ue4umu41}\includegraphics[width=0.4\textwidth]{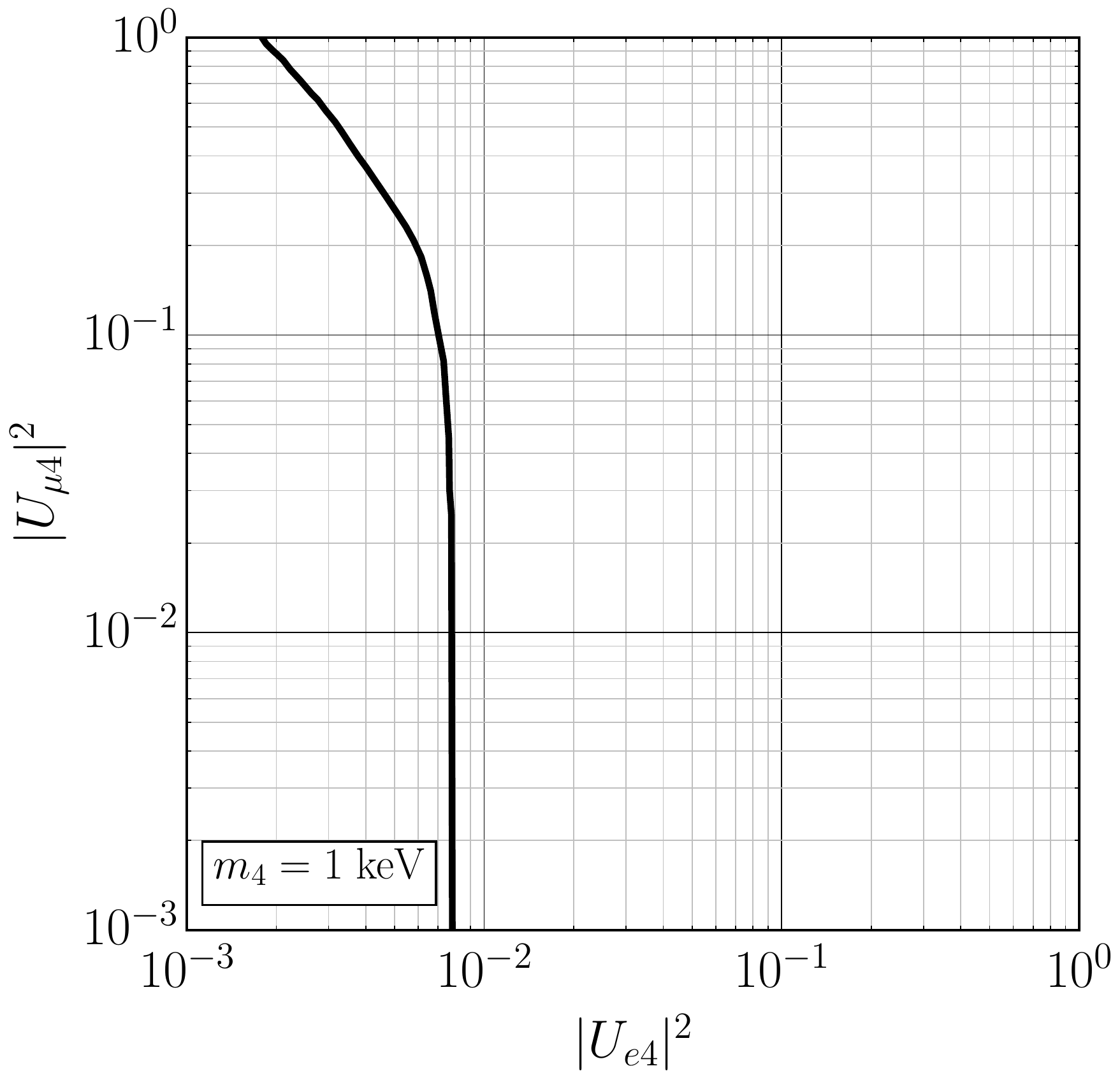}}
\subfigure[]{\label{ue4utau41}\includegraphics[width=0.4\textwidth]{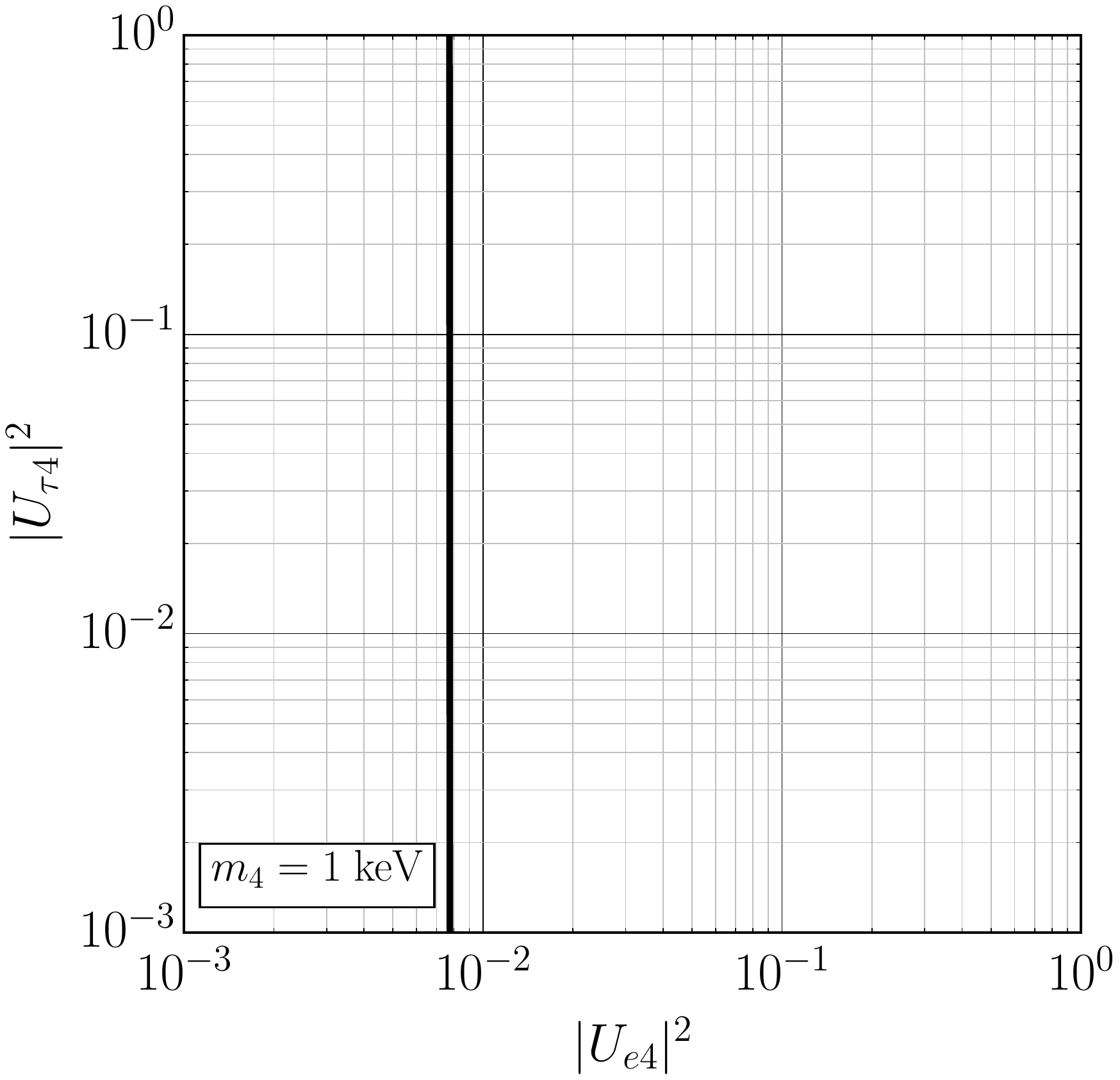}}\\
\subfigure[]{\label{umu4tau41}\includegraphics[width=0.4\textwidth]{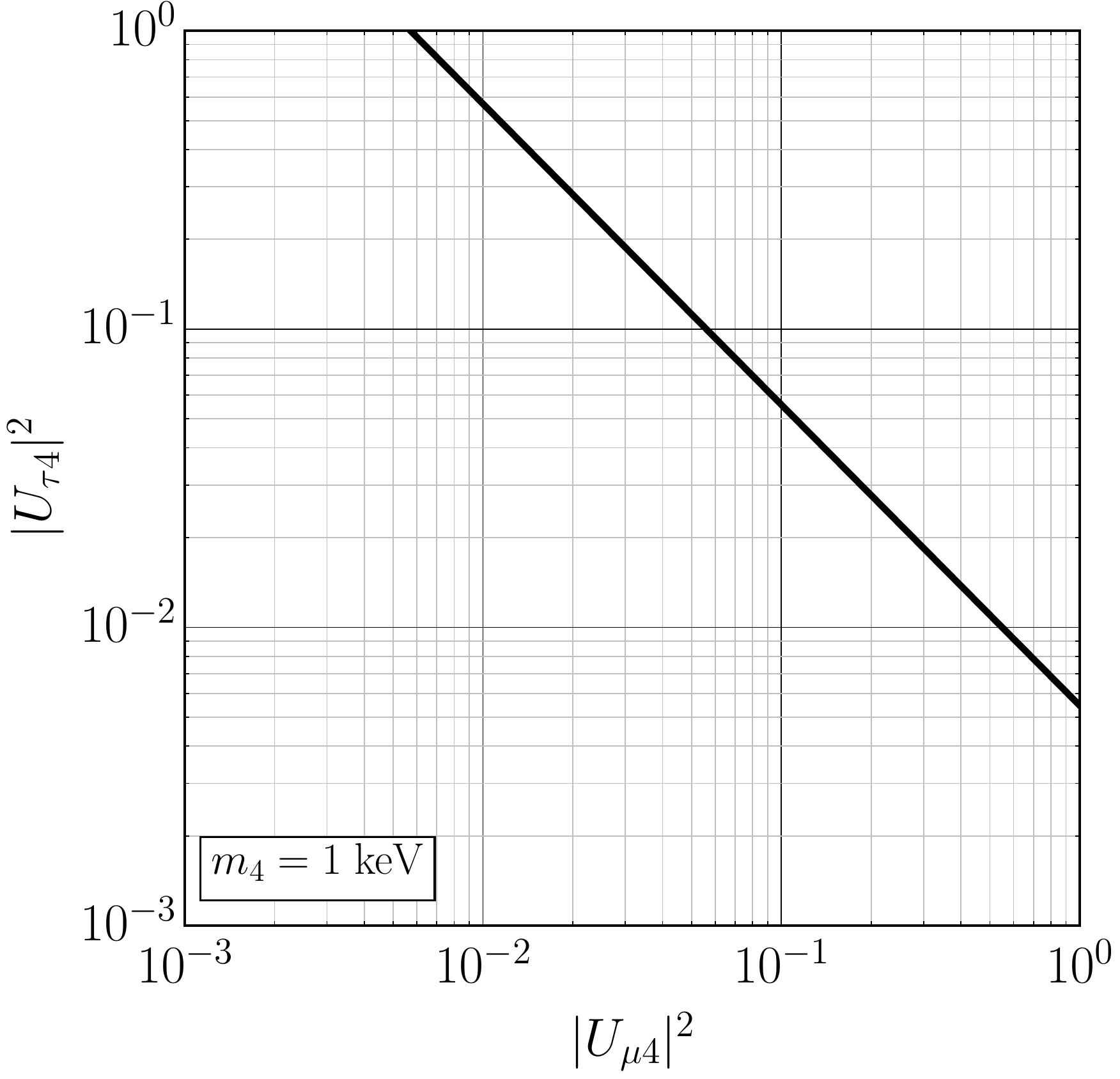}}
\caption{ The two-dimensional 99\% CL upper limits on (a) $|U_{e4}|^2$ vs.~$|U_{\mu4}|^2$, (b) $|U_{e4}|^2$ vs.~$|U_{\tau4}|^2$, and (c) $|U_{\mu4}|^2$ vs.~$|U_{\tau4}|^2$, when $m_4 = 1$ keV, assuming lepton number conservation.   }
\label{offdiagonal_1keV}
\end{figure}

We include Figs.~\ref{offdiagonal_100GeV} and~\ref{offdiagonal_1keV} in order to show two-dimensional marginalized limits on the mixing-matrix elements, for $m_4 = 100$ GeV and $m_4 = 1$ keV, respectively, assuming lepton number conservation.  Here, when $m_4=100$ GeV, the shape of two-dimensional limits on $|U_{e4}|^2$ versus $|U_{\mu4}|^2$ (Fig.~\ref{ue4umu4100}) is dominated by constraints from tests for lepton universality and $\mu-e$ conversion on Ti, while the two-dimensional limits on $|U_{e4}|^2$ versus $|U_{\tau4}|^2$ (Fig.~\ref{ue4utau4100}) and $|U_{\mu4}|^2$ versus $|U_{\tau4}|^2$ (Fig.~\ref{umu4tau4100}) are dominated by only tests for lepton universality.   
When $m_4=1$ keV, the shapes of two-dimensional limits on $|U_{e4}|^2$ versus $|U_{\mu4}|^2$ (Fig.~\ref{ue4umu41}) and $|U_{e4}|^2$ versus $|U_{\tau4}|^2$ (Fig.~\ref{ue4utau41}) are due to constraints from beta decay and neutrino oscillations, while the two-dimensional limits on $|U_{\mu4}|^2$ versus $|U_{\tau4}|^2$ (Fig.~\ref{umu4tau41}) are determined by only neutrino oscillations.  

The limits shown in Fig.~\ref{offdiagonal_1keV} reveal that, given the data under consideration and for light enough $m_4$, it is impossible to place bounds on $|U_{\mu 4}|^2$ and $|U_{\tau 4}|^2$ that are independent from the values of  the other elements of the mixing matrix ($|U_{e 4}|^2$, $|U_{\mu 4}|^2$, and $|U_{\tau 4}|^2$). This is depicted clearly in Figs.~\ref{umu4},~\ref{utau4}, where the 99\% CL limits on $|U_{\mu 4}|^2$ and $|U_{\tau 4}|^2$ are trivial for $m_4\lesssim1$~MeV and $m_4\lesssim200$~MeV, respectively.  Other data, not discussed here, do constrain the heavy neutrino hypothesis even for such light values of $m_4$. Atmospheric neutrino data \cite{Fukuda:1998mi}, for example, reveal that $|U_{\mu 3}|^2\neq 0$, and there is strong evidence -- from atmospheric data \cite{Abe:2012jj} and data from OPERA \cite{Agafonova:2015jxn} -- that $|U_{\tau 3}|^2$ is not zero. Solar neutrino data \cite{Fukuda:2001nj,Ahmad:2002jz}, on the other hand, reveal that  $|U_{\mu 2}|^2\neq 0$ or $|U_{\tau 2}|^2\neq 0$. Since, in the scenario under consideration, $|U_{\alpha 4}|^2=1-\sum_{i=1,2,3}|U_{\alpha i}|^2$, for $\alpha=e,\mu,\tau$, current ``standard'' oscillation data forbid large values of $|U_{\alpha 4}|^2$ for all $\alpha=e,\mu,\tau$. A detailed analysis along these lines is outside the scope of this analysis, but information can be extracted from a careful look at recent studies of the unitarity of the neutrino mixing matrix (see, for example, Refs.~\cite{Qian:2013ora,Parke:2015goa}). Qualitatively, $|U_{\mu 4}|^2$ and $|U_{\tau 4}|^2$ larger than a few tens of percent are excluded for all $m_4$ values larger than several eV. Similarly, data from cosmological surveys should allow one to rule out very large  $|U_{\alpha 4}|^2$ and small $m_4$, even if the heavy neutrino decays invisibly and quickly, since the effective number of neutrinos ($N_{\rm eff}$) is likely to be experimentally distinguishable from SM expectations. However, detailed bounds are model dependent.

Overall, we find that our limits on $|U_{e4}|^2$, $|U_{\mu4}|^2$, and $|U_{\tau4}|^2$ are dominated by dedicated experimental searches to the kinematic signatures of a heavy neutrino, i.e., those discussed in Section~\ref{invconst}.  However, we note that if limits on $\mu-e$ conversion and $Br(\mu\rightarrow 3e)$ are improved by further experimental efforts, nontrivial constraints on $|U_{e4}|^2$ and $|U_{\mu4}|^2$ will emerge when $m_4\gtrsim 1$ GeV, independent of any assumptions regarding $\nu_4$ decay. If a specific model of $\nu_4$ decay renders it invisible to detection, then the limits presented here are applicable. On the other hand, constraints on a model where $\nu_4$ decays visibly can be dramatically different.  For example, if one assumes that $\nu_4$ decays predominately through the weak interactions, many of the constraints from Section~\ref{invconst} would be altered or replaced with those from experiments that directly search for unique decay signatures of the heavy neutrino. This scenario is very strongly constrained (see, for example, Refs.~\cite{Atre:2009rg, Drewes:2015iva, Vincent:2014rja, Abreu:1996pa, Adriani:1992pq, Orloff:2002de, Astier:2001ck, ATLAS:2012ak, Khachatryan:2014dka, Aaij:2014aba, Aaij:2012zr, Deppisch:2015qwa}).  However, constraints using the $\nu_4$ decay products are model specific.  It is for these reasons that we focus not on a specific model of $\nu_4$ decay, but instead augment the decay-independent constraints in Section~\ref{modelindepconstraints} with constraints using the {\it conservative} phenomenological decision that $\nu_4$ decays invisibly.

\begin{acknowledgments}
We are grateful for useful conversations and feedback from 
Rodrigo Alonso, 
Bogdan Dobrescu,
Marco Drewes,
Sean Dobbs,
Patrick Fox, 
Roni Harnik, 
and 
Ruth Van de Water.  
We also thank Marco Drewes for comments on an earlier draft of this manuscript.
The work of AK is supported in part by the DOE grants \#DE-FG02-91ER40684 and \#DE-SC0009919, and AdG is supported in part by the DOE grant \#DE-FG02-91ER40684.  
\end{acknowledgments}

\bibliography{bib}{}

\end{document}